\documentclass{aa}  

\usepackage{graphicx}
\usepackage{mathtools} 
\usepackage{relsize}
\usepackage{multirow}
\usepackage{multicol}
\usepackage{booktabs}
\usepackage{subfigure}
\usepackage[flushleft]{threeparttable}

\usepackage{siunitx}
\usepackage{natbib}
\bibpunct{(}{)}{;}{a}{}{,} 
\usepackage{txfonts}
\usepackage[colorlinks,citecolor=blue]{hyperref}
\begin{document}

   \title{A giant planet shaping the disk around the very \\ low-mass star CIDA 1}

   \author{P. Curone
          \inst{1},
          A. F. Izquierdo
          \inst{2, 3},
          L. Testi
          \inst{2, 4},
          G. Lodato
          \inst{1},
          S. Facchini,
          \inst{1, 2},
          A. Natta
          \inst{5},
          P. Pinilla
          \inst{6},
          N. T. Kurtovic
          \inst{6},\\
          C.~Toci
          \inst{1},
          M.~Benisty
          \inst{7,8},
          M.~Tazzari
          \inst{9},
          F. Borsa
          \inst{10},
          M. Lombardi
          \inst{1},
          C. F. Manara
          \inst{2},\\
          E. Sanchis
          \inst{11,2},
          \and
          L. Ricci
          \inst{12}
          }

   \institute{Dipartimento di Fisica, Università degli Studi di Milano, via Celoria 16, 20133 Milano, Italy\\
              \email{pietro.curone@unimi.it}
         \and
             European Southern Observatory, Karl-Schwarzschild-Str. 2, 85748 Garching bei München, Germany
         \and
             Leiden Observatory, Leiden University, P.O. Box 9513, NL-2300 RA Leiden, The Netherlands
         \and
            INAF – Osservatorio Astrofisico di Arcetri, Largo E. Fermi 5, 50125 Firenze, Italy
         \and
            School of Cosmic Physics, Dublin Institute for Advanced Studies, 31 Fitzwilliams Place, Dublin 2, Ireland
         \and
            Max-Planck-Institut für Astronomie, Königstuhl 17, 69117, Heidelberg, Germany
         \and
            Unidad Mixta Internacional Franco-Chilena de Astronomía, CNRS, UMI 3386. Departamento de Astronomía, Universidad de Chile, Camino El Observatorio 1515, Las Condes, Santiago, Chile
         \and
            Univ. Grenoble Alpes, CNRS, IPAG, 38000 Grenoble, France
         \and
            Institute of Astronomy, University of Cambridge, Madingley Road, CB3 0HA Cambridge, UK
         \and
            INAF – Osservatorio Astronomico di Brera, Via E. Bianchi 46, 23807 Merate (LC), Italy
         \and
            Freie Universit{\"a}t Berlin, Institute of Geological Sciences, Malteserstr. 74-100, 12249 Berlin, Germany
         \and
            Department of Physics and Astronomy, California State University Northridge, 18111 Nordhoff Street, Northridge, CA 91130, USA
             }

   \date{Received 25 November 2021 / Accepted 17 May 2022}

 
  \abstract
   {Exoplanetary research has provided us with exciting discoveries of planets around very low-mass (VLM) stars (\mbox{$0.08\,\mathrm{M}_\odot\lesssim M_\star \lesssim 0.3\,\mathrm{M}_\odot$}; e.g., TRAPPIST-1 and
Proxima Centauri). However, current theoretical models still strive to explain planet formation in these conditions and do not
predict the development of giant planets. Recent high-resolution observations from the Atacama Large Millimeter/submillimeter Array (ALMA)  of the disk around CIDA~1, a VLM star in Taurus, show substructures that hint at the presence of a massive planet.}
   {We aim to reproduce the dust ring of CIDA~1, observed in the dust continuum emission in ALMA Band~7 (\SI{0.9}{\mm}) and Band~4 (\SI{2.1}{\mm}), along with its $^{12}$CO ($J=3{-}2$) and $^{13}$CO ($J=3{-}2$) channel maps, assuming the structures are shaped by the interaction of the disk with a massive planet. We seek to retrieve the mass and position of the putative planet, through a global simulation that assesses planet-disk interactions to quantitatively reproduce protoplanetary disk observations of both dust and gas emission in a self-consistent way.}
   {Using a set of hydrodynamical simulations, we model a protoplanetary disk that hosts an embedded planet with a starting mass of between $0.1$ and $4.0\,\mathrm{M}_{\mathrm{Jup}}$ and initially located at a distance of between 9 and $\SI{11}{\mathrm{au}}$ from the central star. We compute the dust and gas emission using radiative transfer simulations, and, finally, we obtain the synthetic observations, treating the images as the actual ALMA observations.}
   {Our models indicate that a planet with a minimum mass of ${{\sim}}1.4\,\mathrm{M}_{\mathrm{Jup}}$ orbiting at a distance of ${{\sim}} 9{-}\SI{10}{\mathrm{au}}$ can explain the morphology and location of the observed dust ring in Band 7 and Band 4.  We match the flux of the dust emission observation with a dust-to-gas mass ratio in the disk of ${{\sim}}10^{-2}$.  We are able to reproduce the low spectral index (${{\sim}} 2$) observed where the dust ring is detected, with a ${{\sim}} 40{-}50\%$ fraction of optically thick emission.  Assuming a $^{12}$CO abundance of $5\times10^{-5}$ and a $^{13}$CO abundance 70 times lower, our synthetic images reproduce the morphology of the $^{12}$CO ($J=3{-}2$) and $^{13}$CO ($J=3{-}2$) observed channel maps where the cloud absorption allowed a detection. From our simulations, we estimate that a stellar mass  $M_\star=0.2\,\mathrm{M}_{\odot}$ and a systemic velocity $\varv_{\mathrm{sys}}=\SI{6.25}{\km\per\s}$ are needed to reproduce the gas rotation as retrieved from molecular line observations. Applying an empirical relation between planet mass and gap width in the dust, we predict a maximum planet mass of ${{\sim}}4 {-} \SI{8}{\mathrm{M}_{\mathrm{Jup}}}$.}
   {Our results suggest the presence of a massive planet orbiting CIDA~1, thus challenging our understanding of planet formation around VLM stars.}

   \titlerunning{A giant planet around CIDA 1}
   \authorrunning{P. Curone et al.}
   \maketitle
%

\section{Introduction}

Planets have been detected around very low-mass (VLM) stars, defined as stars with masses \mbox{$\lesssim 0.3 \, \mathrm{M}_\odot$} but still above the hydrogen-burning limit of ${{\sim}}0.08 \, \mathrm{M}_\odot$, below which the brown dwarf regime begins \citep{1987ARA&A..25..473L}. Particularly fascinating are the cases of TRAPPIST-1, a  \mbox{${\sim} 0.085 \, \mathrm{M}_\odot$} star in the solar neighborhood with a system of seven rocky planets \citep{2017Natur.542..456G}, and Proxima Centauri, the closest star to the Sun, which has a mass of \mbox{${\sim} 0.12 \, \mathrm{M}_\odot$} and hosts a terrestrial planet, a super-Earth candidate, and a newly discovered sub-Earth candidate (\citealt{2016Natur.536..437A, 2020SciA....6.7467D}, \citealt{2022A&A...658A.115F}). In general, data from the Kepler spacecraft show that the occurrence rate of small planets (with radii of $1.0{-}2.8\,\mathrm{R}_\oplus$) is 3.5 times higher for M~dwarfs than FGK stars \citep{2015ApJ...814..130M}. Even more intriguingly, giant planets have been confirmed orbiting around VLM stars and brown dwarfs. \citet{2005A&A...438L..25C} directly imaged a ${\sim}5\, \mathrm{M}_{\mathrm{Jup}}$ planet around a ${\sim}25\, \mathrm{M}_{\mathrm{Jup}}$ young brown dwarf, and \citet{2019Sci...365.1441M} discovered a planet with a minimum mass of $0.46\, \mathrm{M}_{\mathrm{Jup}}$ orbiting a \mbox{${\sim} 0.12 \, \mathrm{M}_\odot$} star. However, giant planet formation in this low-stellar-mass regime remains a conundrum. 

Planets originate in protoplanetary disks and, generally, the most supported theory to explain their formation is the core accretion model \citep{1996Icar..124...62P}. Assuming this scenario, the main problems for planet formation around VLM stars are the fast dust radial drift \citep{2013A&A...554A..95P} and the apparent lack of material necessary for disks in the low-stellar-mass regime to generate planets \citep{2016A&A...593A.111T, 2020A&A...633A.114S}. Several theoretical studies \citep{2007MNRAS.381.1597P, 2020A&A...638A..88L, 2020MNRAS.491.1998M} performed numerical simulations to assess planet formation around VLM stars and brown dwarfs in a core accretion scenario. They have shown that rocky planet formation in this condition is possible, but the emergence of gas giants is always excluded. A viable explanation for the formation of giant planets around VLM stars seems to be the fragmentation of a disk in its early stages due to gravitational instability \citep{2020A&A...633A.116M}.

Observational evidence of protoplanetary disks around VLM stars and brown dwarfs has been collected for decades via multiple observatories (e.g., \citealt{1998A&A...335..522C, 2001A&A...376L..22N, 2002A&A...393..597N, 2005ApJ...626..498M}). Nowadays, high-resolution and high-sensitivity observations from the Atacama Large Millimeter/submillimeter Array (ALMA) can provide essential information to shed light on planet formation in this low-stellar-mass regime \citep{2014ApJ...791...20R, 2016A&A...593A.111T}. Nonetheless, only a minimal sample of protoplanetary disks in the low-mass regime has been resolved at high resolution. The reason lies in their smaller size and fainter emission compared to disks around more massive stars \citep{2017ApJ...841..116H,2019ApJ...878..103R, 2020A&A...633A.114S}. Hence, they require more demanding observations. 

Recently, \citet{2021A&A...645A.139K} presented the first small survey of disks around VLM stars at high resolution observed by ALMA (${\sim} 0.1''$ at $\SI{0.87}{mm}$, Band 7). The authors selected a sample of six low-mass disks located in the Taurus star-forming region, focusing on the brightest objects to maximize the likelihood of observing substructures. Three sources showed  structures in their dust emission, which could be explained by the interaction with massive planets. Among these disks is CIDA~1 (\object{2MASS J04141760+2806096}), the subject of our work.

This source was first observed by \citet{1993PASP..105..686B}, and the circumstellar disk was detected by \citet{2009ApJ...701..698S} in millimeter wavelengths. Then, with observations from ALMA Cycle 0 at an angular resolution of about $0.4 \arcsec$ (corresponding to ${\sim} \SI{55}{\mathrm{au}}$ at \SI{137.5}{\mathrm{pc}}, the estimated distance to CIDA~1\footnote{Throughout the work, we assume this value as the distance to CIDA~1 for consistency with  \citet{2021A&A...649A.122P}.  Recently, \citet{2021AJ....161..147B} used Gaia EDR3 \citep{2021A&A...649A...1G} and derived a distance of $133.4^{+0.7}_{-0.9}\,\mathrm{pc}$. Such a distance differs by ${\approx}3\%$ from the value in \citet{2021A&A...649A.122P}, implying a systematic difference of the same order in the spatial scales derived from our models and a discrepancy of ${\sim}6\%$ in the estimated value of the disk mass.}), \citet{2014ApJ...791...20R} showed a dust disk and detected the CO ($J=3{-}2$) line emission.

An inner cavity with a radius of ${\sim}$ 20 au in the dust emission was first detected by \citet{2018A&A...615A..95P}, using observations in Band 7 from ALMA Cycle 3 with a resolution of $0.21\arcsec \times 0.12 \arcsec$ (${\sim} 29\times \SI{17}{\mathrm{au}}$). To explain such structure, the authors exploited the \citet{2006Icar..181..587C} criterion for gap opening in the gas, assuming a \citet{1973A&A....24..337S} viscosity value of $10^{-4}$ and a stellar mass $M_\star = 0.1 \, \mathrm{M}_\odot$. They found that a planet with a minimum mass of ${\sim} 0.3 \, \mathrm{M}_{\mathrm{Jup}}$ orbiting at a distance of $\SI{15}{\mathrm{au}}$ from the central star could carve the observed cavity.

\begin{figure}
   \centering
   \includegraphics[width=\hsize]{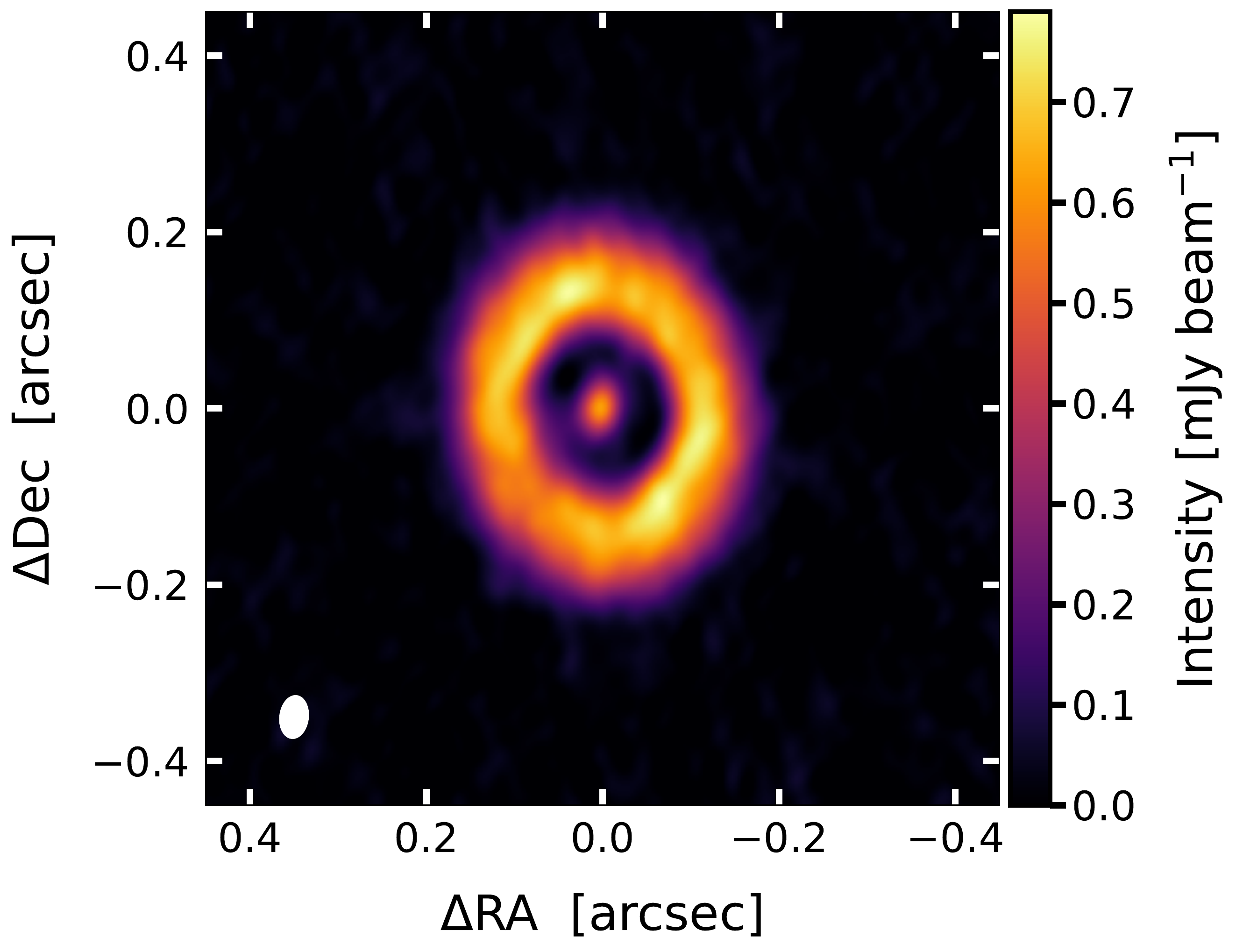}
      \caption{ALMA observation in Band 7 (0.9 mm) of the system CIDA~1 \citep{2021A&A...649A.122P}, with a full width at half maximum (FWHM) beam size  of $0.050\arcsec \times 0.034\arcsec$, indicated by the white ellipse in the bottom-left corner.
              }
         \label{fig:CIDA1_B7_obs}
   \end{figure}

The most recent high-resolution observation of CIDA~1 was presented by \citet{2021A&A...649A.122P}. They showed ALMA Cycle 6 observations (ALMA project 2018.1.00536.S, PI: A. Natta) for dust continuum in Band 7 (0.9 mm) and Band 4 (2.1 mm) at a resolution of ${\sim} 0.050'' \times 0.034''$ (${\sim} 7\times\SI{5}{\mathrm{au}}$). These images clearly reveal a dust ring, whose emission peaks at ${\sim}$ 20 au from the star, surrounding a gap with an unresolved inner disk (see Fig.~\ref{fig:CIDA1_B7_obs}). Furthermore, these observations also detected the gas line emission from $^{12}$CO ($J=3{-}2$) and $^{13}$CO ($J=3{-}2$). The authors used 1D dust evolution simulations and excluded that a planet more massive than $M_\star = 0.5 \, \mathrm{M}_\mathrm{Jup}$ could cause the observed dust gap.

In this work we aim to reproduce the observed dust and gas emission of the disk to evaluate whether the presence of the observed substructures can be explained by the interactions of the disk with an embedded planet. While \citet{2021A&A...649A.122P} assumed an analytical gap shape, we obtain it from hydrodynamical simulations. We employed full 3D modeling to take the dynamical and radiative aspects of the disk into account, along with the effects introduced by interferometric observations with ALMA. Furthermore, in our work we compared the results of our models not only with observed dust emission but also gas emission. We used 3D hydrodynamical simulations to study how planets with different masses influence the dynamics of gas and dust in a disk around a VLM star. After that, we used 3D Monte Carlo radiative transfer simulations to compute the dust continuum emission and the gas line emission. Finally, we treated the output results as real interferometric observations and compared our models to the collected data. The key parameters that we derive are the mass and position of the supposed planet around CIDA~1. The presence of a gas giant planet around a young VLM star would call into question our current theory of planet formation.

This paper is organized as follows: In Sect.~\ref{sect:methods} we describe the numerical simulations and procedures used to obtain the final synthetic observations. In Sect.~\ref{sect:results} we present the results on the modeling of the dust and gas emission, comparing them with the ALMA observations of CIDA~1. Section~\ref{sect:discussion} is devoted to the discussion of our findings: we analyze the simulated  spectral index and optical depth, evaluate whether planet-induced perturbations could be detected in the gas line emission, and constrain the minimum and maximum mass of the putative planet. We also compare our modeling with previous mass estimates of the planet and investigate its possible origin. Finally, in Sect.~\ref{sect:conclusions} we summarize our results.

\section{Methods}
\label{sect:methods}

\subsection{Gas and dust hydrodynamical simulation}

We performed a set of five hydrodynamical simulations with the smoothed particle hydrodynamics (SPH) code \mbox{\textsc{Phantom}} \citep{2018PASA...35...31P}. This code has been extensively used for the modeling of protoplanetary disks, with the intent of reproducing the observed structures in both the dust and gas emission (e.g., \citealp{2015MNRAS.453L..73D}, \citealp{2017MNRAS.464.1449R}, \citealp{2018ApJ...860L..13P}, \citealp{2018MNRAS.477.1270P}, \citealp{2019MNRAS.483.4114C}, \citealp{2019MNRAS.486.4638U}, \citealp{2020MNRAS.499.2015T}, \citealp{2020MNRAS.495.1913V}, \citealp{2020ApJ...888L...4T}). SPH formulation, and the \textsc{Phantom} implementation in particular, includes exact conservation of mass, momentum, and angular momentum, along with a self-consistent computation of the stellar and planet accretion rates and migrations \citep{2018PASA...35...31P}.
We used the one-fluid \citep{2014MNRAS.440.2136L, 2018MNRAS.477.2766B} multigrain \citep{2018MNRAS.476.2186H} algorithm, which discretizes the disk with a single set of SPH particles, each containing information of gas and dust with different grain sizes. This method is suitable for treating dust grains not fully decoupled from the gas. Such a condition is fulfilled if the  midplane Stokes number $\mathrm{St} \propto \rho_0 a_\mathrm{d} / \Sigma_\mathrm{g} < 1$, where $\rho_0$ is the dust grain intrinsic density, $a_\mathrm{d}$ the grain size, and $\Sigma_\mathrm{g}$ the gas surface density.

\subsubsection{Disk model}

We selected the initial parameters for our simulations, listed in Table~\ref{table:Phantom_params}, based on the observations and estimates of   \citet{2018A&A...615A..95P, 2021A&A...649A.122P}. The system consists of a central star surrounded by a disk  composed of gas and dust, sampled by  $N_{\mathrm{SPH}}=10^6$ SPH particles. The disk has a gas mass of $M_\mathrm{gas} = 1.5 \times 10^{-3} \, \mathrm{M}_\odot$ and initially extends from $R_\mathrm{in} = \SI{1}{\,\mathrm{au}}$ to $R_\mathrm{out} = \SI{100}{\,\mathrm{au}}$. Following a common practice (e.g., \citealt{2020MNRAS.499.2015T}), we assumed a power law with an inner and an outer taper for the initial gas surface density profile,
\begin{equation}
        \Sigma_\mathrm{g} (R) = \Sigma_\mathrm{c} \left(\frac{R}{R_\mathrm{in}} \right)^{-p} \exp \left[-\left(\frac{R}{R_\mathrm{c}}\right)^{2-p}  \right] \, \left(1-\sqrt{\frac{R_{\mathrm{in}}}{R}} \,\right) \,,
        \label{eq:sim_surf_dens_prof}
\end{equation} 
where $\Sigma_\mathrm{c}$ is a normalization constant depending on the total disk mass, \mbox{$R_\mathrm{c} = \SI{80}{\,\mathrm{au}}$} is the characteristic radius of the outer exponential taper, and we set $p=1.5$ (as assumed in \citealp{2018A&A...615A..95P}). Our model adopts a locally isothermal equation of state  $P = \rho_{\mathrm{g}} c_\mathrm{s}^2$, with the following sound speed radial profile:
\begin{equation}
        c_\mathrm{s} (R) = c_\mathrm{s,in} \biggl(\frac{R}{R_ {\mathrm{in}}}\biggr)^{-q},
        \label{eq:c_s}
\end{equation}
where $\rho_{\mathrm{g}}$ is the gas volume density, $c_\mathrm{s,in}$ is the sound speed at the inner disk radius and $q=0.3$ (derived from the spectral energy distribution; see Appendix \ref{appendix:q_parameter}).
The gas in the  disk is in vertical hydrostatic equilibrium, which leads to the following aspect ratio:
\begin{equation}
    \frac{H(R)}{R} = \frac{c_\mathrm{s}(R)}{\varv_{\mathrm{K}}(R)} = \frac{H(R_{\mathrm{in}})}{R_{\mathrm{in}}}\,\left( \frac{R}{R_{\mathrm{in}}} \right)^{1/2 - q} \, ,
\end{equation}
where $\varv_{\mathrm{K}}$ is the Keplerian velocity and the aspect ratio at the inner radius is  $H(R_{\mathrm{in}}) / R_{\mathrm{in}} = 0.08$. With this choice of parameters, the mean value of the ratio between the azimuthally averaged SPH smoothing length $\langle h \rangle$ and the scale height $H$ is $\langle h \rangle / H \approx 0.14$, implying that the disk is vertically resolved. We modeled the disk viscosity, which regulates the angular momentum transport, using the SPH artificial viscosity $\alpha_{\mathrm{AV}}$ introduced by \citet{2010MNRAS.405.1212L}. We set the \citet{1973A&A....24..337S} viscosity $\alpha_{\mathrm{SS}} = 5\times10^{-3}$, corresponding to $\alpha_{\mathrm{AV}} = 0.36$. This value directly affects the timescale regulating the disk viscous evolution (higher viscosity generally leading to faster evolution) and on the interaction between the disk and a planet, especially regarding the gas gap carved by a planet (see Sect.~\ref{sect:min_MP}).

   \begin{table}[]
      \caption[]{Initial parameters used for the simulations with \textsc{Phantom}.}
         \label{table:Phantom_params}
         $
         \begin{array}{p{0.3\linewidth}l}
            \hline
            \hline
            \noalign{\smallskip}
            Parameters & \text{Value} \\
            \noalign{\smallskip}
            \hline
            \noalign{\smallskip}
                        $N_{\text{SPH}}$ & 10^6 \\
                        $M_{\star}  [\mathrm{M}_\odot] $ &  $0.2$\\
                        $R_{\star \, \text{acc}} $ [au]         & 1 \\
                        $R_{\text{in}} $ [au]   & 1 \\
                        $R_{\text{out}} $ [au]  & 100 \\
                        $R_{\text{c}} $ [au]    & 80 \\
                        $p$ & 1.5 \\
                        $q$ & 0.3 \\
                        $H(R_{\mathrm{in}}) / R_{\mathrm{in}}$ & 0.08 \\
                        $\alpha_{\text{SS}}$ & 5\times10^{-3} \\
                        $\alpha_{\text{AV}}$ & 0.36 \\
                        $\langle h \rangle / H$ & 0.14 \\
                        $M_{\text{gas}} $ [$\mathrm{M}_{\odot}$]        & 1.5 \times 10^{-3}\\
                        $M_{\text{dust}} / M_{\text{gas}} $     & 10^{-2} \\
                        $a_{\text{d}}$ [cm]& [a_{\text{min}}=1.6\times 10^{-5}, \, a_{\text{max}}= 0.1, N = 11]\\
                        $\xi$ & 3.5 \\
                        $\rho_{\text{0}} \, [\text{g cm}^{-3}]$ & 2 \\
                        \noalign{\smallskip}
                        \hline
                        \noalign{\smallskip}
                        $R_{\text{P} \, \text{acc}} $ [au]      & 1/4 \text{ of Hill radius}\\
                        $d_{\text{P},0} $ [au]  & 10.0 \\
                        $M_{\text{P},0} $ [$\mathrm{M}_{\mathrm{Jup}}$]         & 0.1, 0.5, 1.0, 1.5, 2.0 \\
            \noalign{\smallskip}
            \hline
            \smallskip
         \end{array}
         $
    \small \textbf{Notes.} Each entry is commented on in the main text.
   \end{table}

The dust component is initially distributed as the gas, following the same surface density profile (Eq.~\ref{eq:sim_surf_dens_prof}). The initial dust-to-gas mass ratio in the disk is $10^{-2}$ (the standard value assumed for the interstellar medium, \citealp{1978ApJ...224..132B}), and it is constant throughout the disk. We simulated 11 different grain sizes, logarithmically spaced between $\SI{0.16}{\um}$ and $\SI{1}{\mm}$. The grain size distribution follows the power law  \mbox{$\mathrm{d} N (a_{\mathrm{d}})/ \mathrm{d} a_\mathrm{d} \propto a_\mathrm{d}^{-\xi}$}, truncated at the minimum and maximum grain size, where $a_{\mathrm{d}}$ indicates the grain size and $N(a_{\mathrm{d}})$ is the number of dust grains per grain size. In our simulation, we assumed the typical value $\xi = 3.5$ for the distribution power law \citep{1977ApJ...217..425M}. Simulating various grain sizes allows different levels of coupling between dust and gas to be tested. The intrinsic density of dust grains is $\rho_0 = \SI{2}{\g \, \cm^{-3}}$. 

      \begin{figure*}[]
   \centering
   \includegraphics[width=0.9\textwidth]{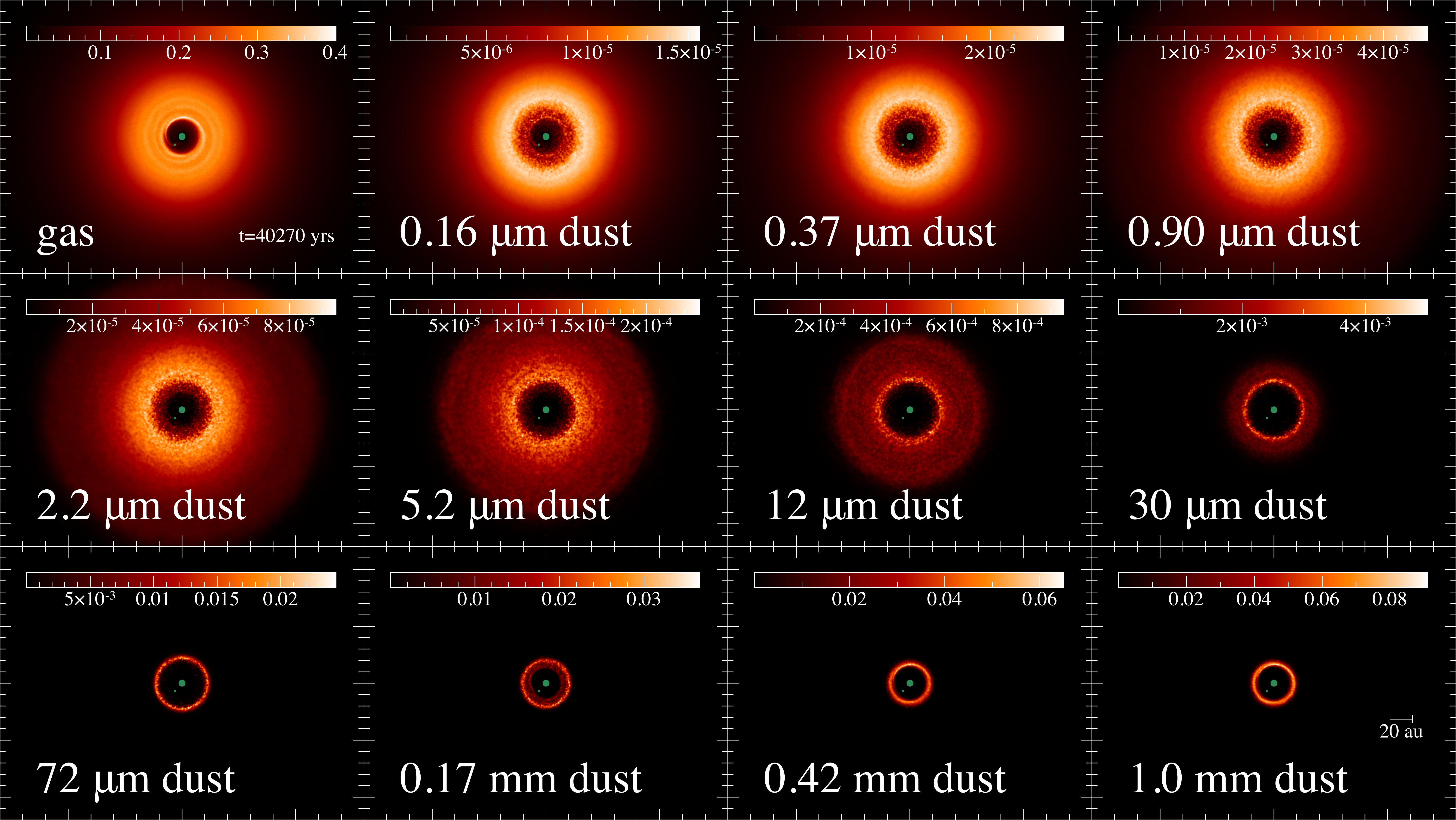}
      \caption{Face-on gas and dust surface density, in units of $\SI{}{\g \, \cm^{-2}}$, from the \textsc{Phantom} simulation with an initial planet mass of $1.5 \, \mathrm{M}_{\mathrm{Jup}}$ and a final planet mass of $2.0 \, \mathrm{M}_{\mathrm{Jup}}$ after ${\approx}\SI{4e4}{\mathrm{yrs}}$ (${\approx} 560$ planet orbits). Gas density is displayed in the top-left corner, and the remaining panels present the surface density for the 11 simulated dust grain sizes. The bigger central green dot represents the star, and the smaller green dot indicates the planet position. Images were produced using the tool \textsc{Splash} \citep{2007PASA...24..159P}.
              }
         \label{fig:phantom_surf_densities}
   \end{figure*}

\subsubsection{Properties of the central star and the embedded planet}

The central star and the planet are modeled as sink particles, which are free to migrate and accrete gas and dust (see Appendix~\ref{appendix:planet_evolution}). The stellar mass of CIDA~1 is known to be in the range ${\sim}0.1-0.2\,\mathrm{M}_\odot$ \citep{2021A&A...649A.122P}. \citet{2021A&A...645A.139K} estimate a mass of ${\sim}0.19\,\mathrm{M}_\odot$ using the \citet{2016ApJ...831..125P} method with the distance of \SI{137.5}{\mathrm{pc}} \citep{2021A&A...649A.122P}. In our simulations, we adopted a stellar mass $M_\star = 0.2 \, \mathrm{M}_\odot$.  The star accretion radius is $R_{\star \, \text{acc}} = \SI{1}{\mathrm{au}}$ and  defines a region where incoming SPH particles are considered accreted onto the star. The planet accretion
radius is chosen to be $R_{\text{P} \, \text{acc}} = 1/4\,R_{\mathrm{H}}$, where $R_{\mathrm{H}}$ is the planet Hill radius, defined as
\begin{equation}
    R_{\mathrm{H}} = \left( \frac{1}{3}\frac{M_{\mathrm{P}}}{M_\star}\right)^{1/3} d_{\mathrm{P}} \, ,
    \label{eq:Hill_radius}
\end{equation}
where $d_{\mathrm{P}}$ is the distance of the planet from the central star.

We performed a set of SPH simulations, each with an embedded planet starting at a distance $d_{\mathrm{P},0} = \SI{10.0}{\mathrm{au}}$ from the star. We varied the initial planet mass $M_{\mathrm{P},0}$: the chosen values are $0.1$, $0.5$, $1.0$, $1.5$, and $2.0$ $\mathrm{M}_{\mathrm{Jup}}$. We let the simulations evolve until they reach a stable configuration. We analyzed the simulation results at $\SI{4e4}{\mathrm{yrs}}$,  corresponding to about 560 planet orbits. To check that no major transients are neglected with this choice, we also computed the dust emission synthetic observations for a longer evolution time of $\SI{8e4}{\mathrm{yrs}}$, equivalent to about 1120 planet orbits.

For illustrative purposes, we show in Fig.~\ref{fig:phantom_surf_densities} the face-on surface densities of the gas and the 11 dust populations from the simulation with initial planet mass $M_{\mathrm{P},0}=1.5\, \mathrm{M}_{\mathrm{Jup}}$, evaluated after $\SI{4e4}{\mathrm{yrs}}$. Here,  the dust radial drift and the dust trap induced by the planet \citep{2012A&A...538A.114P} can be appreciated. This is because the massive planet creates a pressure bump in the gas outside of its orbit, which attracts nearby dust and intercepts dust migrating from the outer regions toward the central star.  The faster motion of larger grains (from ${\sim} 0.1$ to $\SI{1}{\mm}$) results in ring-shaped dust distributions, whereas smaller grains present wider radial profiles of their density, because they have a slower radial drift being more coupled to the gas.

After this first suite of simulations, we aim at better constraining the minimum and maximum planet mass able to reproduce the observed dust emission by running additional simulations. In these cases, the embedded planet is placed at different initial radial distances from the star to find the best match with the peak intensity of the observed ring. The minimum mass of the planet is discussed in Sect.~\ref{sect:min_MP}, where we present the results of two simulations: the first with $M_{\mathrm{P},0}=0.5\, \mathrm{M}_{\mathrm{Jup}}$ and $d_{\mathrm{P},0}=\SI{12.0}{\mathrm{au}}$, the second with $M_{\mathrm{P},0}=1.0\, \mathrm{M}_{\mathrm{Jup}}$ and $d_{\mathrm{P},0}= \SI{11.0}{\mathrm{au}}$. The maximum mass of the planet is assessed in Sect.~\ref{sect:max_MP}, considering a simulation with $M_{\mathrm{P},0}=4.0\, \mathrm{M}_{\mathrm{Jup}}$ and $d_{\mathrm{P},0}=\SI{9.0}{\mathrm{au}}$.

\subsubsection{Limitations of the modeling}
\label{sect:focus_external_disk}

            \begin{figure*}[]
   \centering
   \includegraphics[width=0.9\textwidth]{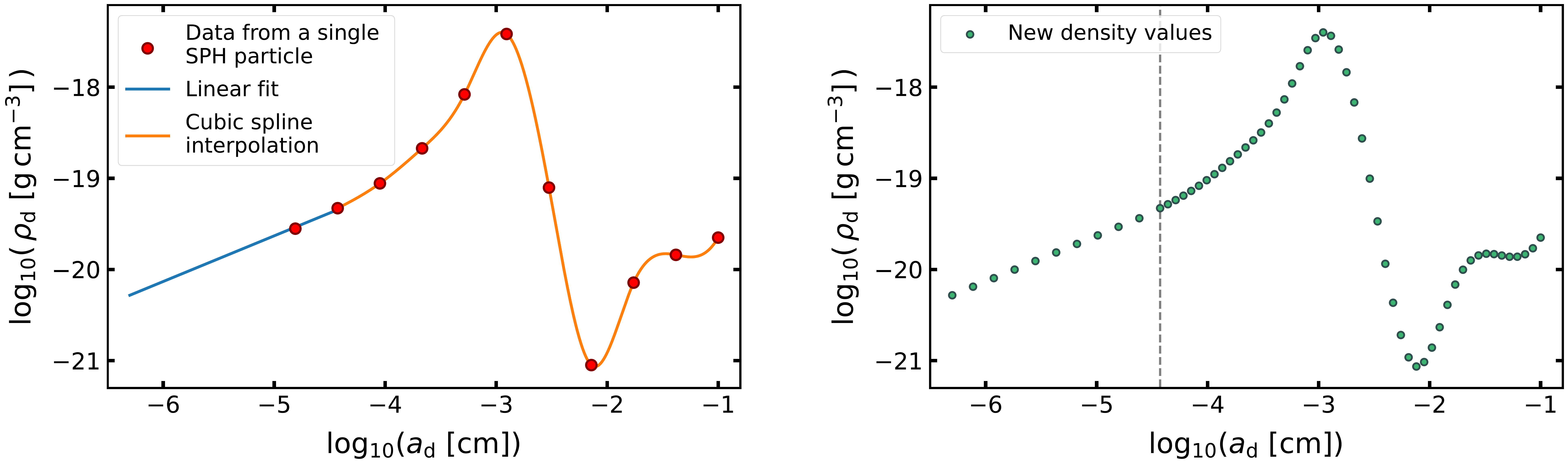}
      \caption{Refinement of the hydrodynamical model for the radiative transfer simulation via the extrapolation and interpolation of the dust densities for a single SPH particle. In the \textit{left panel}, starting from dust density values obtained with SPH simulations (red dots), we perform a linear fit for $\log \rho_{\mathrm{d}}$ values associated with the two smallest simulated grain sizes to obtain densities for grains down to $\SI{5}{\nm}$ (blue line). Then, we use a cubic spline interpolation on the remaining $\log \rho_{\mathrm{d}}$ values (orange line).  In the \textit{ right panel}, with the linear fit, we obtain ten dust densities related to grain sizes between $\SI{5}{\nm}$ and $\SI{0.37}{\um}$ (to the left of the vertical dashed gray line), while with the cubic spline interpolation we obtain 50 densities for grain sizes between $\SI{0.37}{\um}$ and $\SI{1}{\mm}$ (to the right of the vertical dashed gray line).}
        \label{fig:dust_dens_interp}
   \end{figure*}

High-resolution ALMA images of the dust continuum emission from CIDA~1 (Fig.~\ref{fig:CIDA1_B7_obs}) show a dust ring peaking at ${\sim}\SI{20}{\mathrm{au}}$ from the star, a large gap, and an unresolved inner disk at the center, indicating  the presence of dust within a few astronomical units of the star. We focus on modeling the external dust ring, whereas reproducing the inner disk is beyond the aim of this work, given the setup we use for the SPH simulations.

In our models, we adopted $R_{\star \, \mathrm{acc}} = \SI{1}{\mathrm{au}}$; that is, we are not simulating the dynamics of the material within a radius of $\SI{1}{\mathrm{au}}$ from the central star. Adopting a smaller $R_{\mathrm{acc}}$ rapidly causes a substantial increase in computational time due to a higher dynamic range. SPH particles closer to the central star move faster, so smaller time steps and a greater computational cost are needed to simulate their motion. 

Furthermore, an intrinsic consequence of the artificial viscosity implementation in \mbox{\textsc{Phantom}} is that $\alpha_{\mathrm{SS}}\propto\langle h \rangle\propto\rho^{-1/3}$ (in 3D), meaning that the disk viscosity increases wherever the density decreases \citep{2010MNRAS.405.1212L}. A local increase in viscosity leads to a shorter viscous timescale and, therefore, a faster transport of angular momentum in the disk. This description applies to our simulations: if the embedded planet is massive enough to carve a gap in the disk, in that region, the viscosity rises, and this eventually brings to a faster depletion of the inner material accreting onto the central star.

However, these numerical limitations only apply to the inner area, within the first few astronomical units. They do not affect the simulation in outer regions, where the observed dust ring is located and most of the dust flux and mass lie.

\subsection{Radiative transfer}

\subsubsection{Linking \textsc{Phantom} to \textsc{Polaris}}

For the radiative transfer simulations, we employ the 3D Monte Carlo code \textsc{Polaris} \citep{2016A&A...593A..87R}, coupled for the first time with \textsc{Phantom} SPH simulations using the  \textsc{sf}\scalebox{.68}{3}\textsc{dmodels}  code as interface \citep{2018MNRAS.478.2505I}. \textsc{Polaris} is a flexible code that can handle multiple dust compositions, each of them defined by dust material, intrinsic grain density, and grain size distribution between a minimum and maximum grain size. Each dust population is associated with a spatial density distribution. From each SPH simulation, we acquired density distributions for 11 grain sizes, logarithmically spaced between $\SI{0.16}{\um}$ and $\SI{1}{\mm}$. To have a more precise model, we obtained more dust density distributions from this initial data. Considering a single SPH particle, at first, we extrapolated densities for smaller grains. Under the assumption that small dust grains well-coupled to the gas maintain their initial size distribution (i.e., $\mathrm{d} N (a_{\mathrm{d}})/ \mathrm{d} a_\mathrm{d} \propto a_\mathrm{d}^{-3.5}$, which can be rearranged as $\rho_{\mathrm{d}}\propto a_{\mathrm{d}}^{0.5}$, with $\rho_{\mathrm{d}}$ indicating the dust mass density), we fitted  the values of $\log_{10} \rho_{\mathrm{d}}$ associated with the two smallest simulated grain sizes ($\SI{0.16}{\um}$ and $\SI{0.37}{\um}$) using a straight line with slope ${\sim} 0.5$. With this relation, we sampled 10 density values logarithmically spaced between $\SI{5}{\nm}$ and $\SI{0.37}{\um}$. For larger dust grains, we performed a cubic spline interpolation for $\log_{10} \rho_{\mathrm{d}}$ values related to the remaining simulated grain sizes. Thanks to this interpolation, we sampled 50 density values, logarithmically spaced between $\SI{0.37}{\um}$ and $\SI{1}{\mm}$. This procedure is illustrated in Fig.~\ref{fig:dust_dens_interp}. We repeated  this operation for every SPH particle and obtained in total 60 dust density spatial distributions, 10 for grain sizes between $\SI{5}{\nm}$ and $\SI{0.37}{\um}$ and 50 related to grain sizes between $\SI{0.37}{\um}$ and $\SI{1}{\mm}$.

While in SPH the information is contained in point particles, \textsc{Polaris} needs data to be sampled on a 3D grid for the radiative transfer simulations. To maintain the spatial distribution of physical properties as close as possible to the simulation output we used a Voronoi grid. Given a distribution of $N$ points, a Voronoi grid is a partition of space into $N$ convex polygonal cells, each containing exactly one generating point such that any portion of space inside a specific cell is closer to its generating point than to any other \citep{Okabe00}. To avoid numerical artifacts where fewer SPH particles are present, as in the case of gaps and cavities in disks, we introduced a set of ‘‘dummy'' grid points (as in \citealt{2021MNRAS.500.5268I}), which are artificially added particles to which we assigned a negligible density value ($\SI{e-42}{\g.\cm^{-3}}$). We randomly distributed $1/5\,N_{\mathrm{SPH}}$ dummy points; then, we accepted only those located in low-density areas, less sampled by SPH particles. After that, we used the SPH particles and the accepted dummy points as generating points for the Voronoi grid.

\subsubsection{\textsc{Polaris} simulations}

With the 60 dust distributions sampled on a Voronoi grid obtained from the initial SPH models, we then performed the radiative transfer simulations with \textsc{Polaris}. We considered the same dust composition adopted in \citet{2010A&A...512A..15R}. Here, dust grains are spheres composed of astronomical silicates (10\% in volume, optical properties from \citealp{2003ApJ...598.1026D}), carbonaceous materials (20\%, \citealp{1996MNRAS.282.1321Z}), water ice (30\%, \citealp{2008JGRD..11314220W}), and a porosity of 40\%.

First, with \textsc{Polaris}, we computed the dust temperature using a Monte Carlo approach.  From the estimates of \citet{2021A&A...649A.122P}, CIDA~1 has a luminosity $L_\star = 0.15\pm0.03\, \mathrm{L}_{\sun}$. In our simulations, we modeled the radiation as $10^8$ photon packages emitted by the central star, which is assumed as a black body with a surface temperature $T_{\mathrm{eff}}=\SI{3050}{\K}$ and a radius $R_{\star} = 1.3\,\mathrm{R}_{\sun}$, whose total luminosity equals the observed value. Second, the dust continuum emission and the gas line emission are computed via ray-tracing. We fixed the distance of CIDA~1 at \SI{137.5}{\mathrm{pc}}. From the analysis of \citet{2021A&A...649A.122P},  we used an inclination angle $i=37.5^{\circ}$ and a position angle $\mathrm{PA}=11^{\circ}$ for the disk.

The synthetic images are computed at the same wavelengths observed by ALMA. We simulated the dust continuum emission in Band 7 (\SI{0.9}{\mm}) and Band 4 (\SI{2.1}{\mm}). The gas line emission is calculated for $^{12}$CO and $^{13}$CO in the $J=3{-}2$ transition. For the radiative transfer simulation of the gas emission, we assumed that the system is in local thermodynamic equilibrium and that the gas temperature is the same as that of the dust. Works by \citet{2012A&A...541A..91B}, \citet{2013A&A...559A..46B}, and \citet{2018A&A...612A.104F} show that gas and dust in protoplanetary disks in certain conditions may not be fully thermally coupled. However, given the uncertainty in this context, mainly due to a strong dependence on disk turbulence and dust grain size distribution, we made the assumption that gas and dust temperature are coupled and do not introduce further parameters in our simulations. The systemic velocity in our models is fixed at $\SI{6.25}{\km\per\s}$, because this value led to the best match with the spatial distribution of the observed channel maps. We adopted a gas turbulent velocity of $\SI{100}{\m\per\s}$. We included freeze-out and photo-dissociation for CO following the parameterizations  in \citet{2014ApJ...788...59W}. The condition $\mathrm{T}_{\mathrm{gas}} < \SI{20}{\K}$ defines the freeze-out region \citep{2007prpl.conf..751B}. CO is assumed photo-dissociated by UV radiation from the central star and the interstellar radiation field where the gas column density (calculated in the vertical direction) is lower than a critical value $N_{\mathrm{dissoc}} = \SI{1.3e21}{\cm^{-2}}$ \citep{2009A&A...503..323V}. Therefore, in the warm molecular layer between the frozen-out midplane and the photo-dissociated regions at higher altitudes, we set the $^{12}$CO abundance relative to $\mathrm{H}_2$  $[^{12}\mathrm{CO}/\mathrm{H}_2]=5\times10^{-5}$, and the isotopologue abundance ratio $[^{12}\mathrm{CO}/^{13}\mathrm{CO}] = 70$ \citep{1994ARA&A..32..191W}; elsewhere, the CO abundance is zero. 

To match the velocity resolution of the observed gas channel maps, namely \SI{0.5}{\km\per\s} for $^{12}$CO and \SI{1.0}{\km\per\s} for $^{13}$CO, we first computed the simulated channel maps at a higher velocity resolution of \SI{0.045}{\km\per\s}. Then, to obtain the final spectrally convoluted synthetic channels, we averaged 11 contiguous simulated $^{12}$CO channels, centered at the same velocities of the observed channels. We repeated the same procedure for the computed  $^{13}$CO channels, averaging  22 channels to match the velocity resolution of the observations.

\subsection{Synthetic observations}

To obtain the final beam-convoluted synthetic images from our simulations, we treated the full-resolution dust continuum images and gas channel maps computed by \textsc{Polaris} as the actual ALMA observations. Initially, we employed the task \texttt{sampleImage} from the package \textsc{galario} \citep{2018MNRAS.476.4527T} to compute the synthetic visibilities of the  dust continuum images from our models at the same $(u,\varv)$ points of the ALMA observations, in Band 7 and Band 4, respectively. After that, we calculated the residuals between the observations and the simulations. Finally, we performed the imaging and acquired the synthetic observations using the \texttt{tclean} algorithm from the software \textsc{casa}, version 5.7 \citep{2007ASPC..376..127M}. We applied the same parameters used to obtain the dust continuum Band 7 and Band 4 images in \citet{2021A&A...649A.122P}: the Briggs weighting scheme and a robust parameter of 0.5.

At this point, we can compare the flux in our synthetic images to the one from the observations. We consider the total dust mass of our models as a free parameter that can be rescaled by a constant factor in the radiative transfer simulations to match the flux emitted in Band 7 (\SI{0.9}{\mm}) from the observed external dust ring (see Appendix~\ref{appendix:dust_mass_rescaling}). It is possible to apply this procedure as long as the dust-to-gas mass ratio remains~$\ll 1$, so that the dust back-reaction onto the gas is negligible. In our models, this condition is always fulfilled.

We performed the same method to obtain the $^{12}$CO and $^{13}$CO synthetic channel maps, working channel by channel. In the cleaning process, we used a Briggs robust parameter of 1 and applied a uv-tapering with a Gaussian of $0.035\arcsec$ to recover the same beam size of the observations.

\section{Results}
\label{sect:results}

\begin{table}[]
    \caption{Resulting parameters for our simulations with different initial planet masses.}
        \def\arraystretch{1.2}
        \begin{tabular}{c c  c  c  c  c }
                \midrule
                \midrule
                $M_{\text{P},0}$& $t_{\mathrm{Sim}}$ & $M_{\text{P}} $  & $d_{\text{P}} $ & $M_{\text{dust}} $      & $M_{\text{dust}} / M_{\text{gas}}  $\\
                
                [$\mathrm{M}_{\mathrm{Jup}}$] & [yrs] & [$\mathrm{M}_{\mathrm{Jup}}$] & [au] & [$\mathrm{M}_{\odot}$]  &  \\
                
                \midrule
                \multirow{2}{*}{$0.1$} &$ 4\times 10^4$  & 0.2  & 9.4 & $6.0\times 10^{-6}$ &   $6.3\times 10^{-3}$ \\
                &$ 8\times 10^4$  & 0.2  & 9.2 & $4.3\times 10^{-6}$ &   $5.6\times 10^{-3}$ \\
                
                \midrule
                \multirow{2}{*}{$0.5$} &$ 4\times 10^4$  & 0.9  & 9.3 & $4.8\times 10^{-6}$ &   $5.8\times 10^{-3}$ \\
                &$ 8\times 10^4$  & 0.9  & 9.4 & $3.3\times 10^{-6}$ &   $5.1\times 10^{-3}$ \\
                
                \midrule
                \multirow{2}{*}{$1.0$} &$ 4\times 10^4$  & 1.4  & 9.4 & $7.0\times 10^{-6}$ &   $8.7\times 10^{-3}$ \\
                &$ 8\times 10^4$ & 1.5  & 9.4 & $4.0\times 10^{-6}$ &   $6.4\times 10^{-3}$ \\

                \midrule        
                \multirow{2}{*}{$1.5$} &$ 4\times 10^4$  & 2.0  & 9.4 & $7.0\times 10^{-6}$ &   $8.9\times 10^{-3}$ \\
                &$ 8\times 10^4$  & 2.0  & 9.5 & $5.4\times 10^{-6}$ &   $8.8\times 10^{-3}$ \\
                
                \midrule        
                \multirow{2}{*}{$ 2.0 $} &$ 4\times 10^4$  & 2.5  & 9.4 & $7.3\times 10^{-6}$ &   $9.4\times 10^{-3}$ \\
                &$ 8\times 10^4$  & 2.6  & 9.6 & $6.4\times 10^{-6}$ &  $1.1\times 10^{-2}$ \\
                \midrule
        \vspace{-6pt}
        \end{tabular}
        \label{tab:evolution_SPH_params}
      \small \textbf{Notes.} Planet masses are evaluated at two different times during the simulation ($t_{\mathrm{Sim}}$), namely \SI{4e4}{\mathrm{yrs}} and \SI{8e4}{\mathrm{yrs}}, corresponding to about 560 and 1120 orbits of the embedded planet. $M_{\mathrm{P},0}$ indicates the initial planet mass, while $M_{\mathrm{P}}$ and $d_{\mathrm{P}}$ are the planet mass and distance from the central star at the considered $t_{\mathrm{Sim}}$. All planets started at $d_{\text{P},0} = \SI{10.0}{\mathrm{au}}$.
\end{table}

   \begin{figure*}[]
   \centering
   \includegraphics[width=0.76\textwidth]{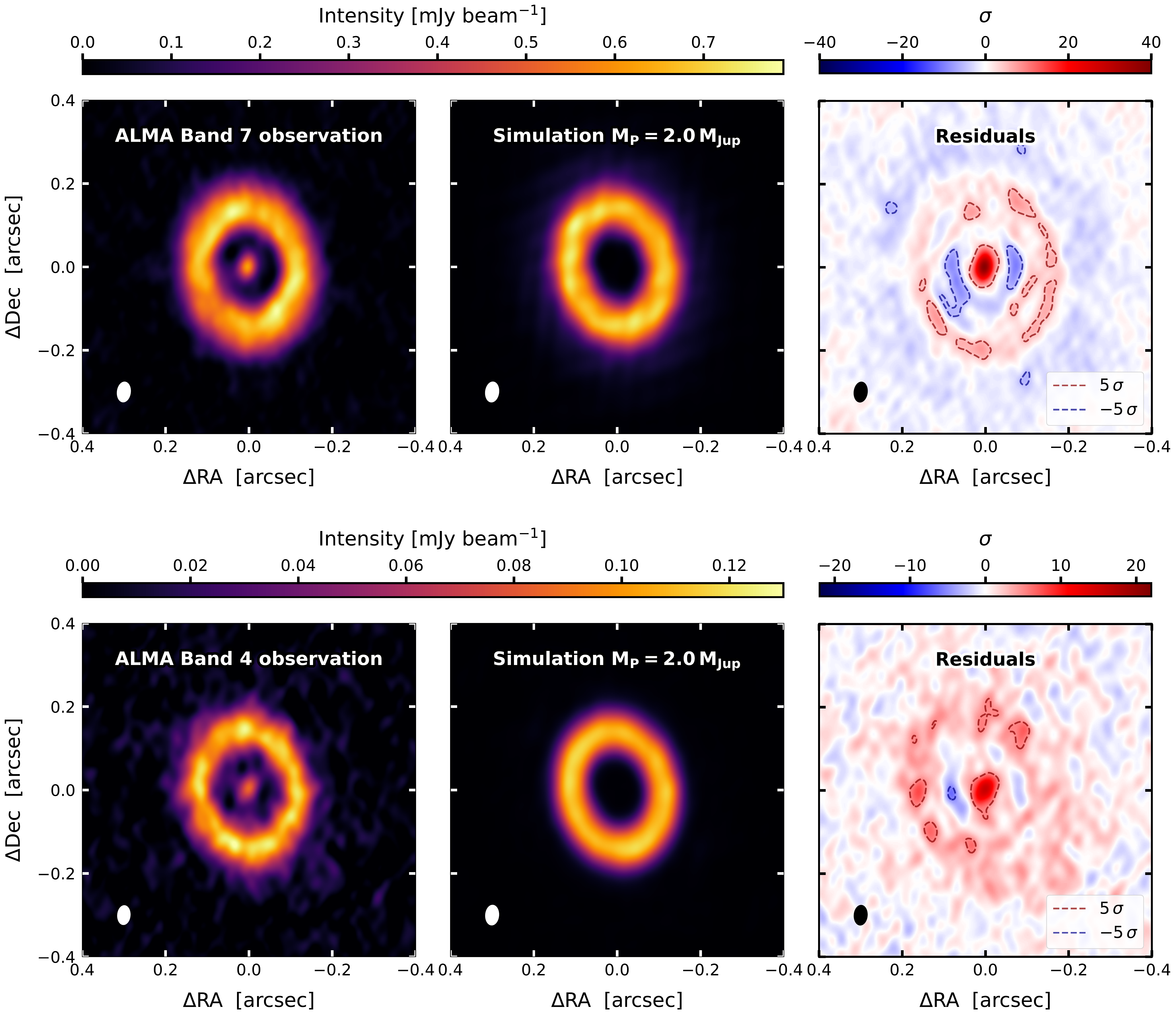}
      \caption{Band 7 (\SI{0.9}{\mm}, \textit{top panels}) and Band 4 (\SI{2.1}{\mm,} \textit{bottom panels}) comparison between  dust continuum images from the ALMA observation and our synthetic observations obtained from the simulation with a final planet mass of $2.0 \, \mathrm{M}_{\mathrm{Jup}}$ after $\SI{4e4}{\mathrm{yrs}}$, along with their residuals. The synthesized beam ($0.050\arcsec \times 0.034\arcsec$ for Band~7, $0.048\arcsec \times 0.032\arcsec$ for Band~4, FWHMs) is shown in the bottom-left corner of each image. The contours in the residuals represent the $5\sigma$ and $-5\sigma$ flux levels, respectively, for Band~7 and Band~4.
              }
         \label{fig:obs_mod_res_2Mpl}
   \end{figure*}

\begin{figure*}[]
   \centering
   \includegraphics[width=0.70\textwidth]{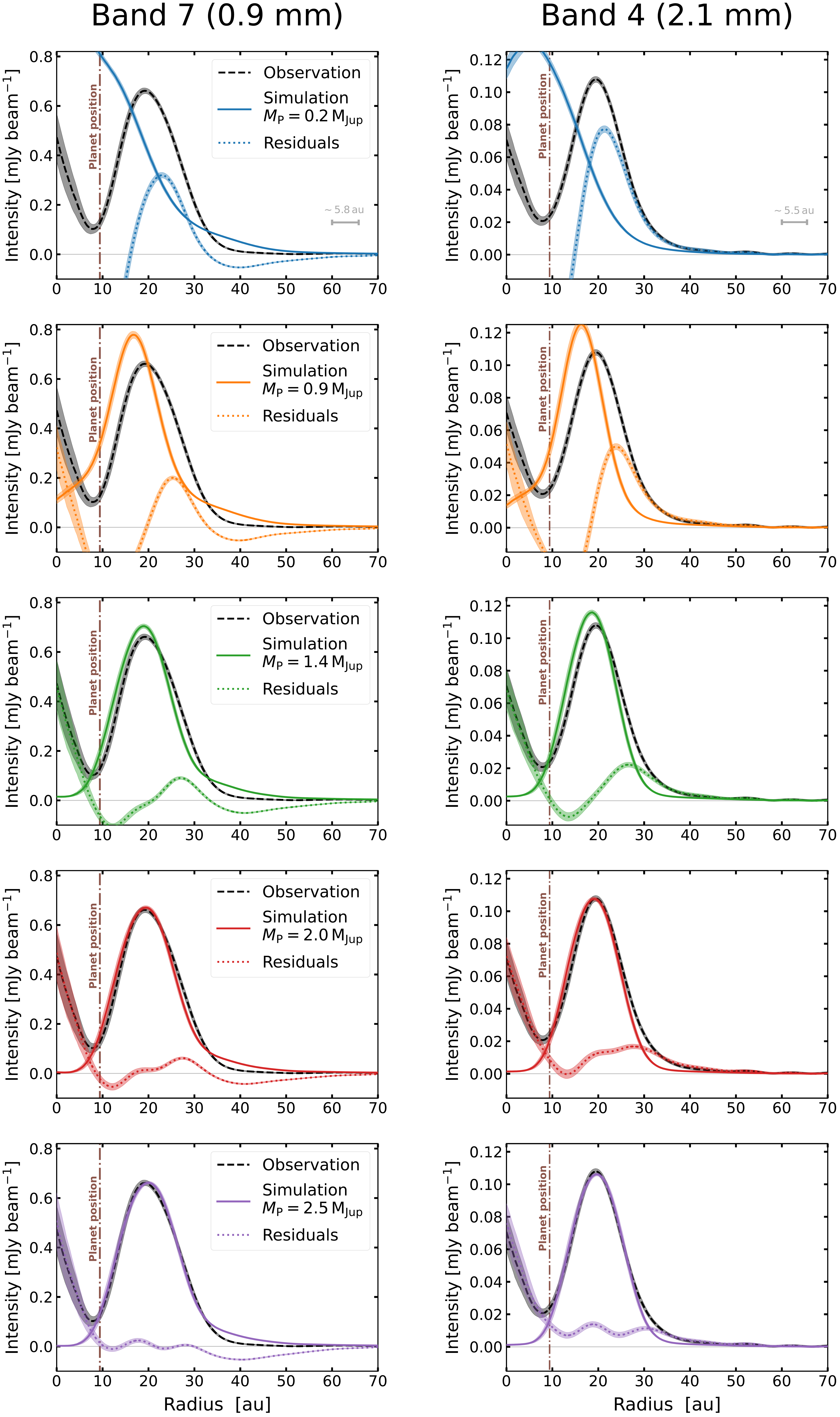}
      \caption{Comparison between radial profiles of azimuthally averaged intensities from the observations, simulations, and corresponding residuals, after the simulations have evolved for $\SI{4e4}{\mathrm{yrs}}$. \textit{The left-hand panels} show results in Band 7, while the right-hand panels refer to Band 4. Each row contains the intensity profiles from the same simulation with a specific final planet mass, ranging from $0.2\,\mathrm{M}_{\mathrm{Jup}}$ (\textit{top row}) to $2.5\,\mathrm{M}_{\mathrm{Jup}}$ (\textit{bottom row}). The distance of the planet from the central star is indicated by the vertical dash-dotted brown line. The mean FWHMs of the synthesized beam in Band~7 and Band~4 are shown as  horizontal bars in the \textit{top row panels}. The radial intensity profiles are obtained by deprojecting the disk, using the disk inclination angle  $i=37.5^{\circ}$, and dividing it into overlapping annuli, whose widths correspond to the averaged synthesized beam diameter. For each bin, uncertainties are calculated by taking the intensity standard deviation for each pixel in the annulus and dividing it by the square root of the number of synthesized beams contained in the area of the annulus.}
         \label{fig:mod&res_profiles}
   \end{figure*}
   
   \begin{figure*}[]
    \centering
    \begin{subfigure}
        \centering
        \includegraphics[width=0.845\textwidth]{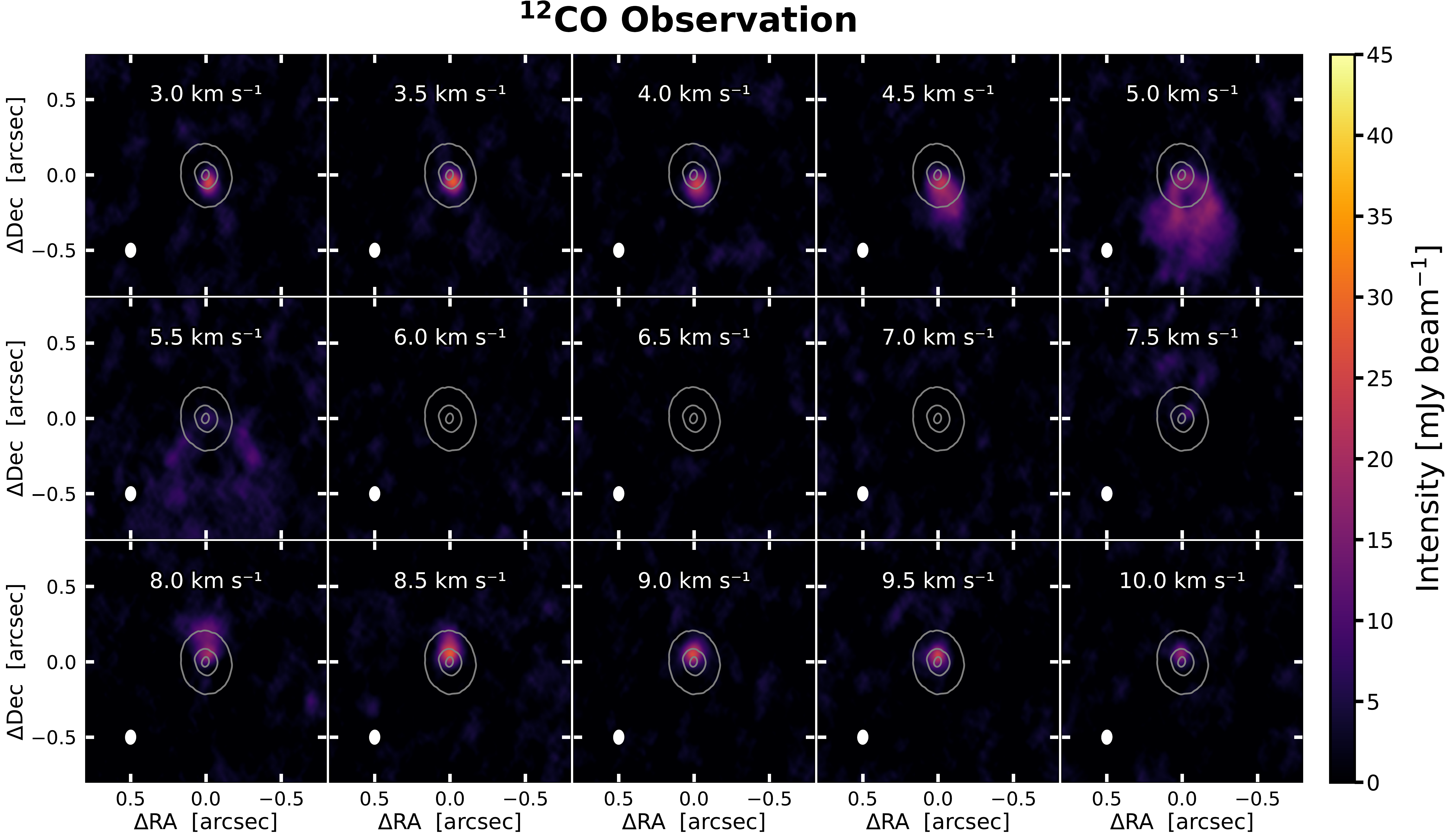}
    \end{subfigure} \\
    \vspace{0.6cm}
    \begin{subfigure}
        \centering
        \includegraphics[width=0.845\textwidth]{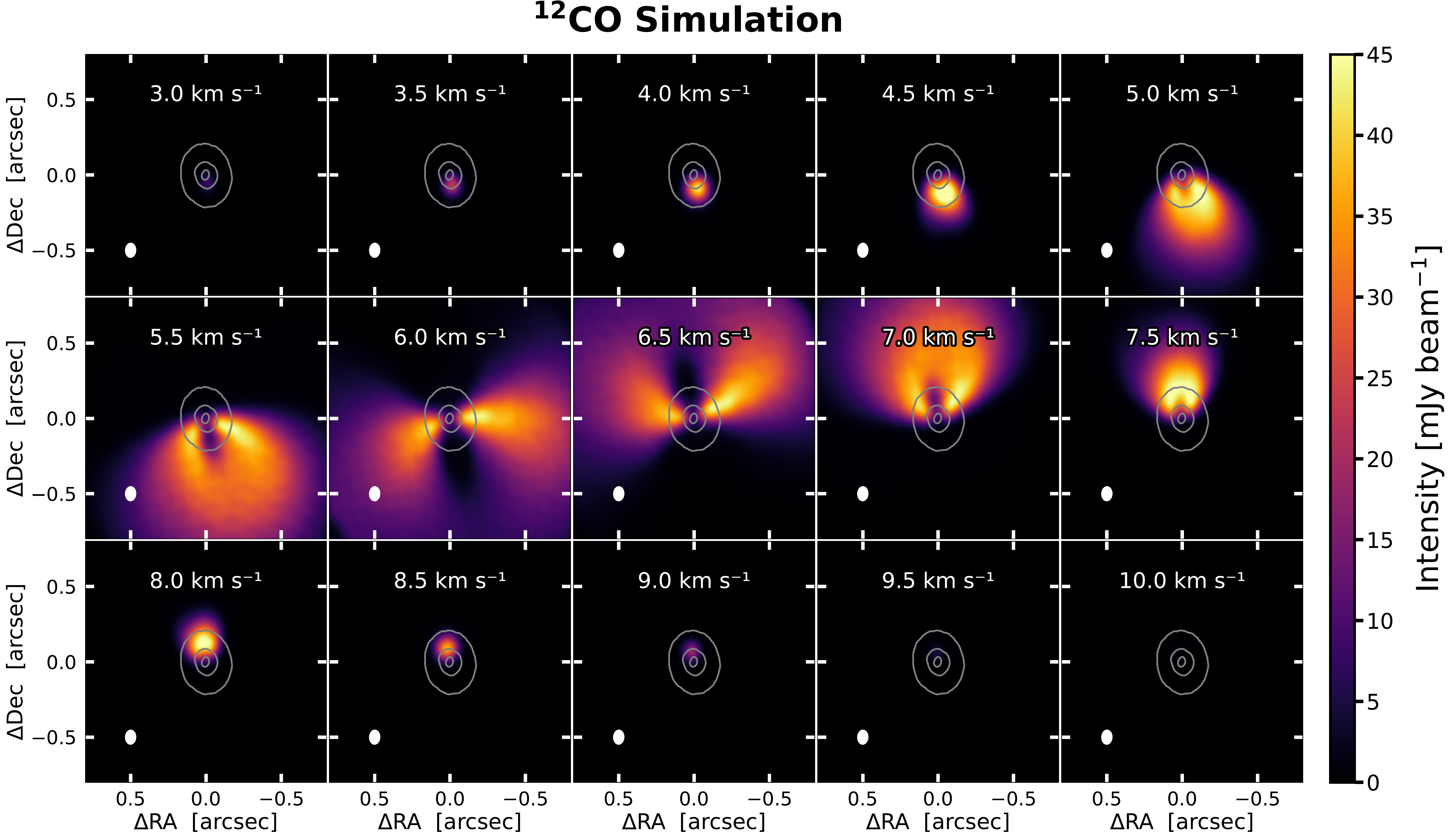}
    \end{subfigure} 
    \caption{Comparison between the ALMA observation (\textit{top panels}) and the synthetic images (\textit{bottom panels}) of CIDA~1 $^{12}$CO ($J=3{-}2$) channel maps. Color scales are the same in the \textit{top and bottom panels}. Synthetic images are obtained from the simulation with a final planet mass of $2.0 \; \mathrm{M}_{\mathrm{Jup}}$ after an evolution time of $\SI{4e4}{\mathrm{yrs}}$. We fixed the systemic velocity at \SI{6.25}{\km\per\s}. The contour level traces the 15$\sigma$ emission from the Band~7 continuum image. The velocity resolution is \SI{0.5}{\km\per\s}, and the central velocities of each channel are indicated on the top of each panel. The synthesized beam ($0.101\arcsec \times 0.075\arcsec$, FWHM) is shown in the lower-left corner of each panel.}
    \label{fig:obs&sim_channels_12CO}
    \end{figure*}
    
       \begin{figure*}[]
   \centering
   \includegraphics[width=0.95\textwidth]{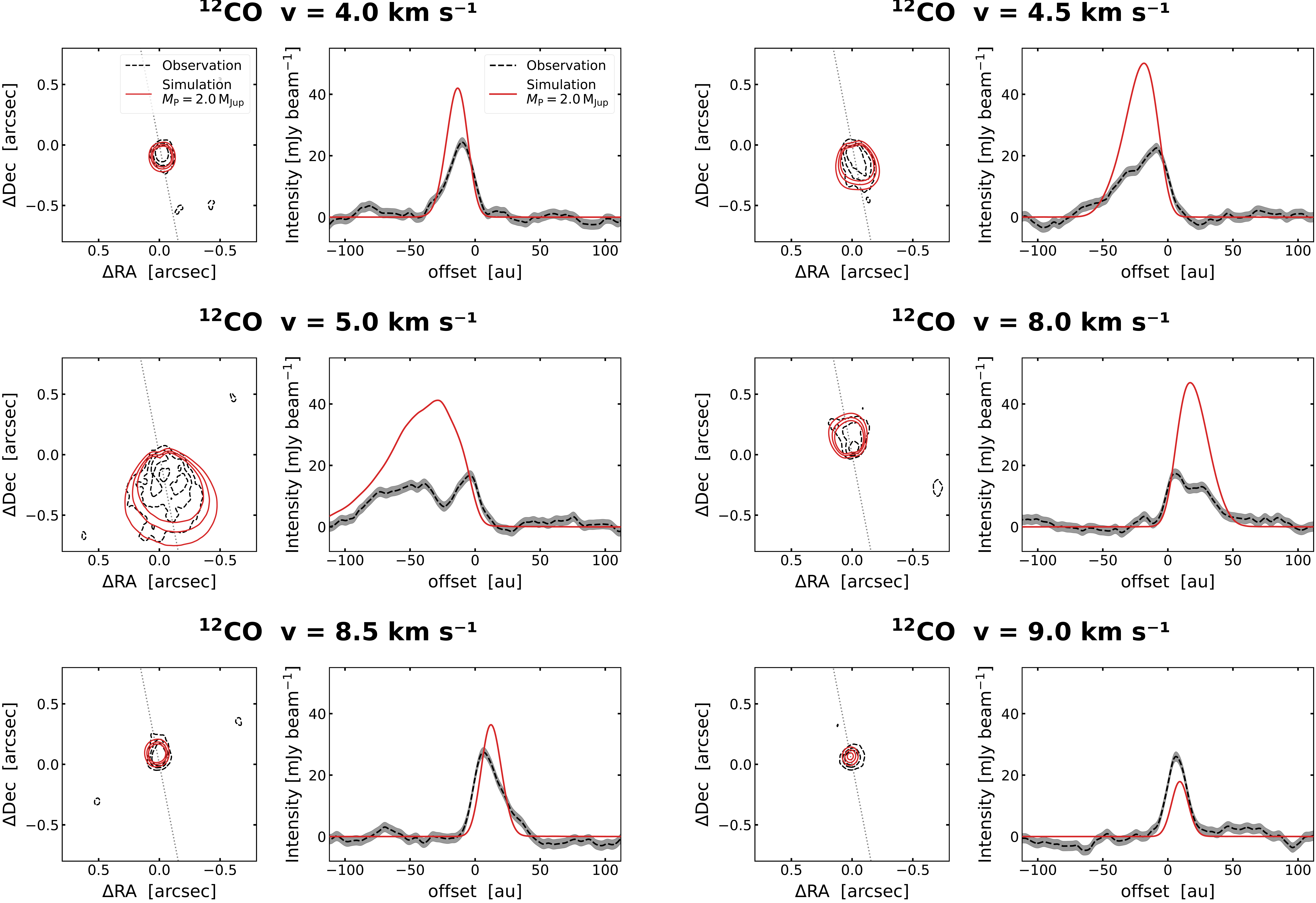}
      \caption{Contour levels (\textit{left panel} in each subplot) at $5,\,10,\,\SI{15}{\mathrm{mJy}\,\mathrm{beam}^{-1}}$, and intensity profile comparisons (\textit{right panel} in each subplot) between the $^{12}$CO observation and simulation with a $2.0 \; \mathrm{M}_{\mathrm{Jup}}$ planet mass after $\SI{4e4}{\mathrm{yrs}}$, in the six channels   not completely obscured by the cloud ($\varv=4.0,\,4.5,\,5.0,\,8.0,\,8.5,\,\SI{9.0}{\km\per\s}$). Intensity profiles are taken along the $\mathrm{PA}=11^{\circ}$ of the disk, and the direction is indicated by the dotted gray line in the contour levels plots. The uncertainty in the observation intensity profile is the $1\sigma$ noise level ($\SI{1.6}{\mathrm{mJy}\,\mathrm{beam}^{-1}}$).}
         \label{fig:12CO_channels_comparison}
   \end{figure*}
   
          \begin{figure*}[]
    \centering
    \begin{subfigure}
        \centering
        \includegraphics[width=0.57\textwidth]{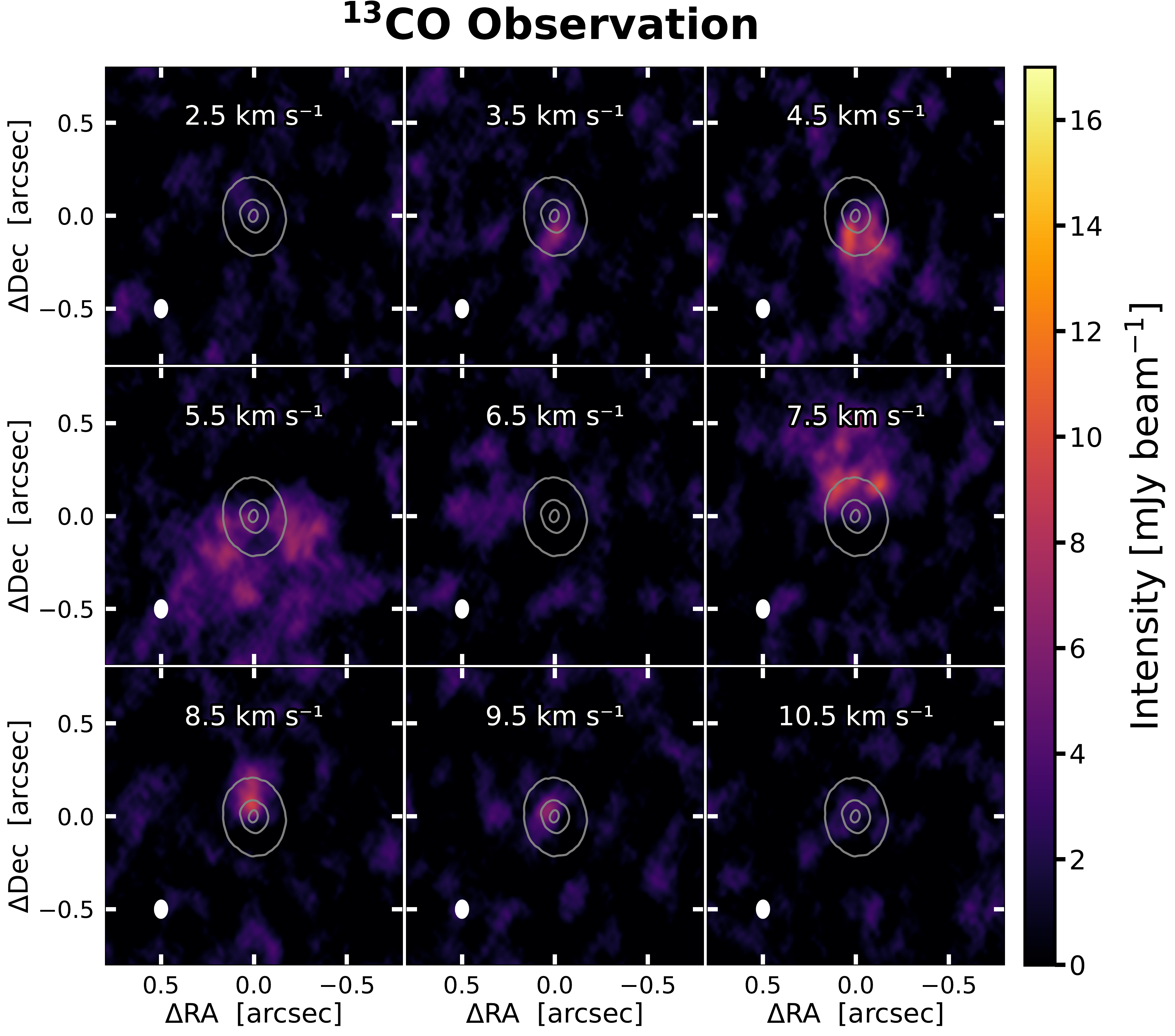}
    \end{subfigure} \\
    \vspace{0.6cm}
    \begin{subfigure}
        \centering
        \includegraphics[width=0.57\textwidth]{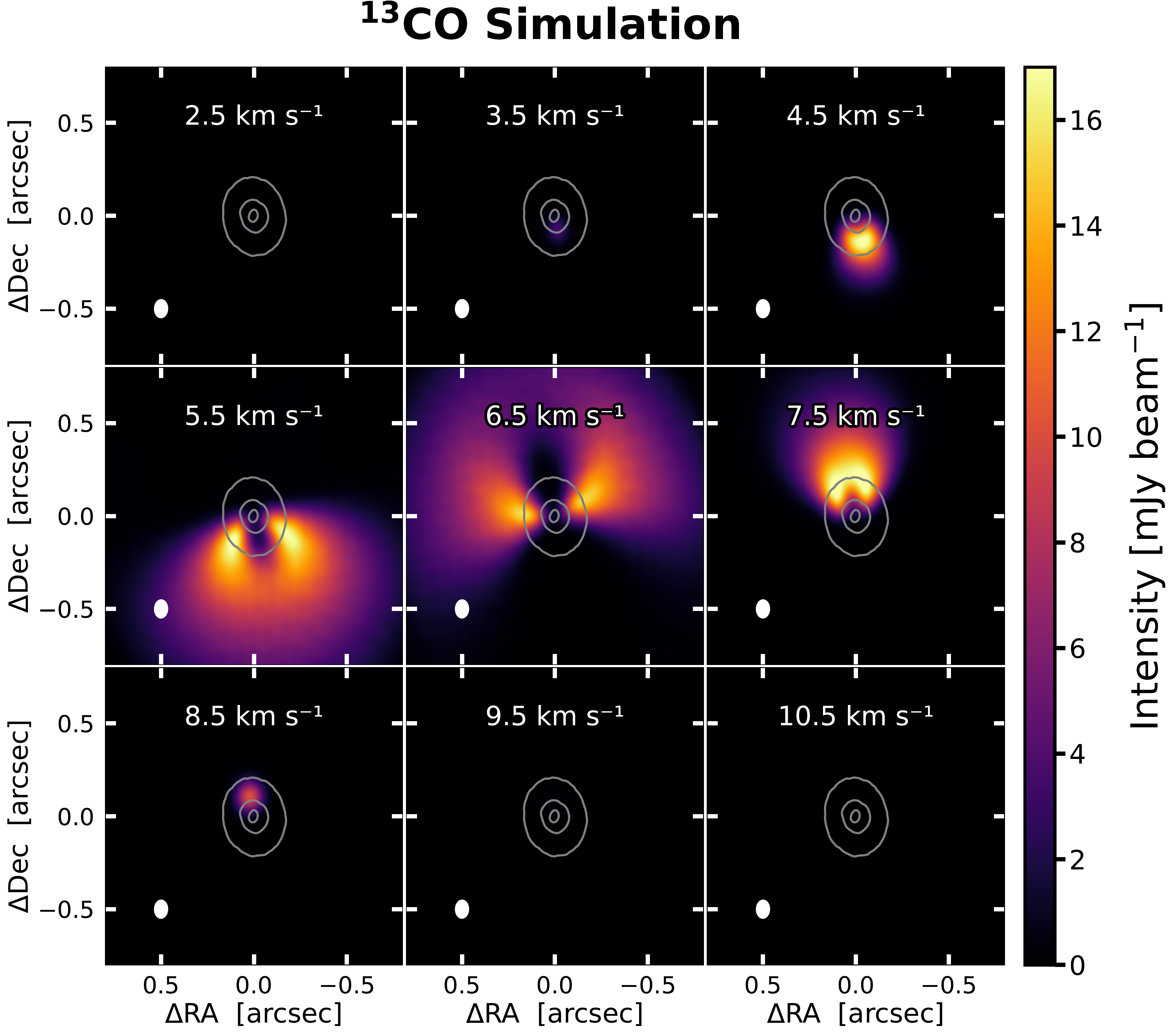}
    \end{subfigure} 
    \caption{Comparison between the ALMA observation (\textit{top panels}) and the synthetic images (\textit{bottom panels}) of CIDA~1 $^{13}$CO ($J=3{-}2$) channel maps. Color scales are the same in the \textit{top and bottom panels}. Synthetic images are obtained from the simulation with a final planet mass of $2.0 \; \mathrm{M}_{\mathrm{Jup}}$ after an evolution time of $\SI{4e4}{\mathrm{yrs}}$. We fixed the systemic velocity at \SI{6.25}{\km\per\s}. The contour level traces the 15$\sigma$ emission from the Band~7 continuum image. The velocity resolution is \SI{1}{\km\per\s}, and the central velocities of each channel are indicated on the top of each panel. The synthesized beam ($0.104\arcsec \times 0.076\arcsec$, FWHM) is shown in the lower-left corner of each panel.}
    \label{fig:obs&sim_channels_13CO}
    \end{figure*}
   
   \begin{figure*}[]
   \centering
   \includegraphics[width=0.95\textwidth]{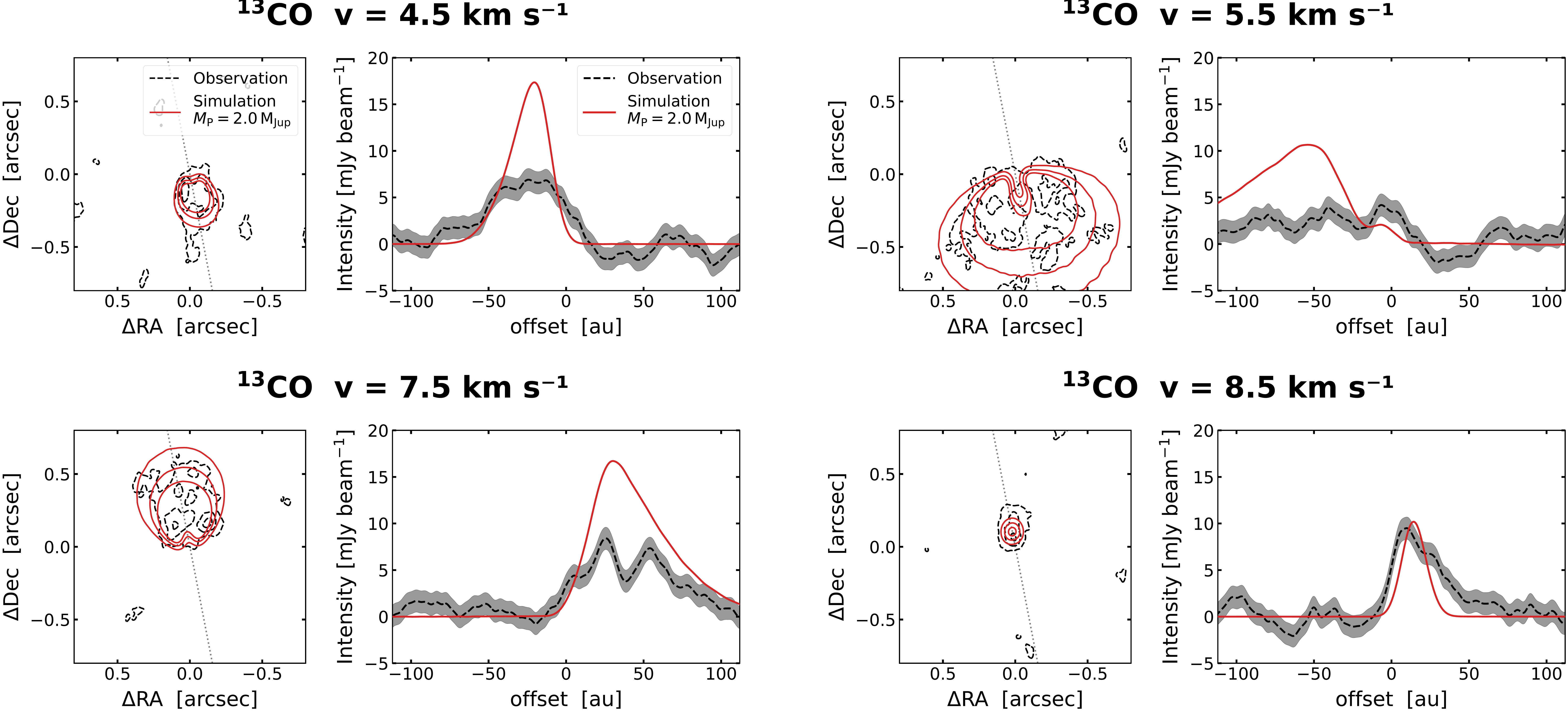}
      \caption{Contour levels (\textit{left panel} in each subplot) at $3,\,6,\,\SI{9}{\mathrm{mJy}\,\mathrm{beam}^{-1}}$, and intensity profile comparisons (\textit{right panel} in each subplot) between the $^{13}$CO observation and simulation with a $2.0 \; \mathrm{M}_{\mathrm{Jup}}$ planet mass after $\SI{4e4}{\mathrm{yrs}}$, in the four channels not completely obscured by the cloud   ($\varv=4.5,\,5.5,\,7.5,\,\SI{8.5}{\km\per\s}$). Intensity profiles are taken along the $\mathrm{PA}=11^{\circ}$ of the disk, and the direction is indicated by the dotted gray line in the contour levels plots. The uncertainty in the observation intensity profile is the $1\sigma$ noise level ($\SI{1.2}{\mathrm{mJy}\,\mathrm{beam}^{-1}}$).}
         \label{fig:13CO_channels_comparison}
   \end{figure*}

The resulting parameters of our models are listed in Table~\ref{tab:evolution_SPH_params}, for each simulation with different initial planet mass  and for the two considered  times $t_{\mathrm{Sim}} = \SI{4e4}{\mathrm{yrs}}$ and $t_{\mathrm{Sim}} = \SI{8e4}{\mathrm{yrs}}$. We note little difference in planet mass ($M_{\mathrm{P}}$) and distance from the central star ($d_{\mathrm{P}}$) in each simulation at the two times, meaning that $\SI{4e4}{\mathrm{yrs}}$ are sufficient for the planet to reach a slowly evolving configuration. Such a finding is confirmed by the plots in Appendix~\ref{appendix:planet_evolution} showing the time evolution for the planet-star distance and the planet mass in the case of the simulation with an initial planet mass of $1.5 \; \mathrm{M}_{\mathrm{Jup}}$. Here, we note a fast transient in the first \SI{e4}{\mathrm{yrs}}, when the planet is carving the gap, followed by a more stable evolution. At this later stage, the system has adapted from the arbitrary initial conditions and reached a quasi-equilibrium state. If the planet is massive enough, such a situation corresponds to the case in which the planet has already carved an internal cavity both in the gas and in the dust, as in the case illustrated in  Fig.~\ref{fig:phantom_surf_densities}.

\subsection{Dust continuum emission}

The first aim of our work is reproducing the dust thermal emission of the dust ring in CIDA~1 observed by ALMA in Band~7 (\SI{0.9}{\mm}) and Band~4 (\SI{2.1}{\mm}), and reported in \citet{2021A&A...649A.122P}. We present in Fig.~\ref{fig:obs_mod_res_2Mpl} the comparison at both wavelengths between the dust continuum image from the ALMA observation and the synthetic image from our simulation with a final planet mass of $2.0\,\mathrm{M}_{\mathrm{Jup}}$ after $\SI{4e4}{\mathrm{yrs}}$, together with their residuals. In the case of this particular simulation, we note that our model is able to generally reproduce the morphology and the intensity of the observed emission from the ring. As expected, our synthetic images do not present the inner disk, which persists in the residuals. Moreover, we note that the residuals in both bands show some areas, located in the region of the ring and the gap, above the $5\sigma$ value or below the  $-5\sigma$ value. Slight shifts in the simulated image centroids could not eliminate these differences. Such inequalities between our model and the observations might be caused by an imperfect estimate of the disk inclination and position angle that we assumed in the simulations or by some actual non-axisymmetric features in the observations.

A thorough comparison between the dust continuum emission from the observations and all our set of simulations after $\SI{4e4}{\mathrm{yrs}}$ is presented in Fig.~\ref{fig:mod&res_profiles}. We show the azimuthally averaged radial profiles of the intensity in the two bands from the observations, all our simulations, and their residuals.

The smallest simulated planet, with a final mass of $0.2\,\mathrm{M}_{\mathrm{Jup}}$, does not produce any major effect on the disk morphology. The $0.9\,\mathrm{M}_{\mathrm{Jup}}$ planet is massive enough to carve a central cavity, but the intensity peak is significantly shifted inward compared to the peak in the observation profiles. The third model, with a planet mass of $1.4\,\mathrm{M}_{\mathrm{Jup}}$, shows a good match with the observation, except for a slight offset in the location of the ring. The two remaining models, hosting the most massive planets with a mass of $2.0$ and $2.5\,\mathrm{M}_{\mathrm{Jup}}$, respectively, best reproduce the observed emission in both Band~7 and Band~4. Looking at Table~\ref{tab:evolution_SPH_params} and considering the results after an evolution time of $\SI{4e4}{\mathrm{yrs}}$, we note that these last three cases also share consistent estimates of the total dust mass needed to match the observed flux, ${\sim}7\times10^{-6}\,\mathrm{M}_\odot$, with a dust-to-gas mass ratio of ${\sim}9\times10^{-3}$.

Results for the dust emission from the simulations after a longer evolution time of \SI{8e4}{\mathrm{yrs}} are presented in Appendix~\ref{appendix:dust_emission_profiles_8e4yrs}. The synthetic profiles for simulations with a minimum planet mass of $2.0\,\mathrm{M}_{\mathrm{Jup}}$ still show a good agreement with the observations in Band~7 (which we take as reference; Appendix~\ref{appendix:dust_mass_rescaling}). However, the simulated profiles in Band~4 have a slightly lower peak intensity compared to the observed ones because the bigger millimeter-sized dust grains, which emit the most at this wavelength and have the highest radial drift, accreted more onto the central star.  Nonetheless, the evident similarities between dust emission after \SI{4e4}{} and \SI{8e4}{\mathrm{yrs}} prove that, at these times, the disks in our simulations have already overcome the initial fast transient phase and are in a dynamically slowly evolving condition, in the absence of dust evolution.

\subsection{Gas line emission}
\label{sect:gas_line_emission}

The dust continuum modeling shown in Fig.~\ref{fig:mod&res_profiles} proves that, while obtaining a good match in the case of a $1.4\,\mathrm{M}_{\mathrm{Jup}}$ planet, the minimum planet mass able to best replicate the observed dust emission is $2.0\,\mathrm{M}_{\mathrm{Jup}}$. Therefore, we decided to use the model with such a planet to compute the $^{12}$CO ($J=3{-}2$) and $^{13}$CO ($J=3{-}2$) channel maps.

In Fig.~\ref{fig:obs&sim_channels_12CO} we compare the $^{12}$CO channel maps observed by ALMA to our synthetic gas emission images. In our model, we set a systemic velocity of $\SI{6.25}{\km\per\s}$. Strong cloud absorption is evident in the central channels of the observation, from ${\sim} 5.5$ to $\SI{7.5}{\km\per\s}$. We analyze the cloud contamination effects in Appendix~\ref{appendix:cloud_abs}, using previous observations of the gas emission from the Taurus molecular cloud by \citet{2010MNRAS.405..759D}.

Our simulation reproduces appropriately the spatial extent of the channels centered at the velocities \mbox{$\varv=4.0,\,4.5,\,5.0,\,8.0,\,8.5,\,\SI{9.0}{\km\per\s}$}. This is more evident in Fig.~\ref{fig:12CO_channels_comparison}, which shows for these channels the comparison between the observed and simulated $^{12}$CO emission in terms of intensity contour levels at $5,\,10,\,\SI{15}{\mathrm{mJy}\,\mathrm{beam}^{-1}}$ and intensity profiles along the $\mathrm{PA}=11^{\circ}$. The channels  $\varv=4.5,\,5.0,\,\SI{8.0}{\km\per\s}$ are characterized by a significant difference in intensity values between the observation and the model. This effect can be explained by a still-present influence of the cloud on these channels, which are just outside the most absorbed velocities. Intensities are more similar in the channels $\varv=4.0,\,8.5,\,\SI{9.0}{\km\per\s}$, where the cloud contamination in the observation is weaker, and the emission comes from regions of the disk well sampled by our simulation.

The observed channel maps also show gas emission from the innermost part of the disk, near the inner disk detected in the dust continuum. This material is also the cause of the central emission in high-velocity channels, such as those at $\varv=\SI{3.5}{\km\per\s}$ and $\varv=9.5{-}\SI{10.0}{\km\per\s}$. As we explain in Sect. \ref{sect:focus_external_disk}, our model does not aim at describing this inner region.

We follow the same procedure for the analysis of the $^{13}$CO emission. Figure~\ref{fig:obs&sim_channels_13CO} presents the comparison between the $^{13}$CO channel maps observed by ALMA and the synthetic ones computed from our simulation. Here, we have a lower velocity resolution of $\SI{1.0}{\km\per\s}$. In the observations, we note that the $\SI{6.5}{\km\per\s}$ channel is completely affected by cloud absorption, which also impacts the channel at $\varv=\SI{5.5}{\km\per\s}$.  In Fig.~\ref{fig:13CO_channels_comparison}, we compare the contour levels at $3,\,6,\,\SI{9}{\mathrm{mJy}\,\mathrm{beam}^{-1}}$ and the intensity profiles at $\mathrm{PA}=11^{\circ}$ from the observed and simulated $^{13}$CO emission. Even in this case, our models are able to reproduce the spatial distribution of the observed emission while presenting higher intensity values, except for the $\SI{8.5}{\km\per\s}$ channel. In particular, we note a strong cloud absorption for the observed intensity profile of the $\SI{5.5}{\km\per\s}$ channel, but the corresponding contour plot shows a good agreement in the extent of the emission between observation and simulation. As for the $^{12}$CO, there is some observed inner emission for the high-velocity channels at $3.5$ and $\SI{9.5}{\km\per\s}$, which cannot be reproduced by our model due to its empty inner cavity. 

From all of these results, we realize that our single model with a final planet mass of $2.0\,\mathrm{M}_{\mathrm{Jup}}$ is able to globally reproduce both the dust and gas emission observed in CIDA~1. In particular, this case perfectly recovers the external dust ring as observed both in Band~7 and Band~4, along with replicating the spatial distribution of the gas emission as traced by $^{12}$CO ($J=3{-}2$) and $^{13}$CO ($J=3{-}2$).


\section{Discussion}
\label{sect:discussion}

\subsection{Stellar mass and systemic velocity}

Our models allow fundamental information regarding the CIDA~1 system to  be recovered. In the observations, the $^{12}$CO and $^{13}$CO  channel maps are strongly affected by cloud absorption in the channels  near the systemic velocity (Sect.~\ref{sect:gas_line_emission} and Appendix~\ref{appendix:cloud_abs}). Hence, we compare our synthetic images to the observations only in those channels probing higher velocities in the disk (Fig.~\ref{fig:12CO_channels_comparison} and~\ref{fig:13CO_channels_comparison}). Models are consistent with observations, being able to reproduce the  spatial distribution of the emission. The stellar mass $M_\star=0.2\,\mathrm{M}_{\odot}$ and systemic velocity $\varv_{\mathrm{sys}}=\SI{6.25}{\km\per\s}$ assumed in our simulations are therefore reasonable estimates of these properties in CIDA~1. This  stellar mass is compatible with the known range ${\sim} 0.1-0.2\,\mathrm{M}_{\odot}$ \citep{2021A&A...649A.122P} and is in very good agreement with the estimate of  $0.19\,\mathrm{M}_{\odot}$ by \citet{2021A&A...645A.139K}. We also simulated the case with $M_\star = 0.1 \, \mathrm{M}_\odot$, but the morphology of the resulting gas channel maps differed significantly from the observed ones; therefore, we decided not to include these models in the simulation sample.

\subsection{Spectral index and optical depth}

\begin{table}[]
      \caption[]{Fluxes in Band 7 and Band 4 and the resulting spatially integrated spectral index for the ALMA observation (accounting only for flux from the external dust ring) and our model from the simulation with a final planet mass of $2.0 \; \mathrm{M}_{\mathrm{Jup}}$ after $\SI{4e4}{\mathrm{yrs}}$.}
        \label{table:spectral_index_spatially_integrated}
        \def\arraystretch{1.2}
        \begin{tabular}{c  c  c  }
                \midrule
                \midrule
                & Observation & Simulation  \\
                & (external ring) &  $M_{\mathrm{P}} = 2.0 \,\mathrm{M}_{\mathrm{Jup}}$ \\
                \midrule
                $F_{B7}$ [mJy] & $29.0\pm2.9$  & $28.6\pm2.9$ \\
                $F_{B4}$ [mJy] & $4.7\pm0.5$ &    $4.0\pm0.4$ \\
                Spectral index $\alpha_{\mathrm{B7},\,\mathrm{B4}}$& $2.08\pm0.16$  & $2.26\pm0.16$ \\
                \midrule
                \vspace{-6pt}
        \end{tabular}
      \small \textbf{Notes.} The areas considered when calculating the spectral index are the same used for the dust mass rescaling process (Appendix~\ref{appendix:dust_mass_rescaling}). The uncertainties for the observations are obtained as the root sum square of the observed noise ($\sigma$) and a 10\% flux error due to calibration. The uncertainties for our simulation only include the 10\% flux error.
\end{table}
   
      \begin{figure*}[]
   \centering
   \includegraphics[width=1\textwidth]{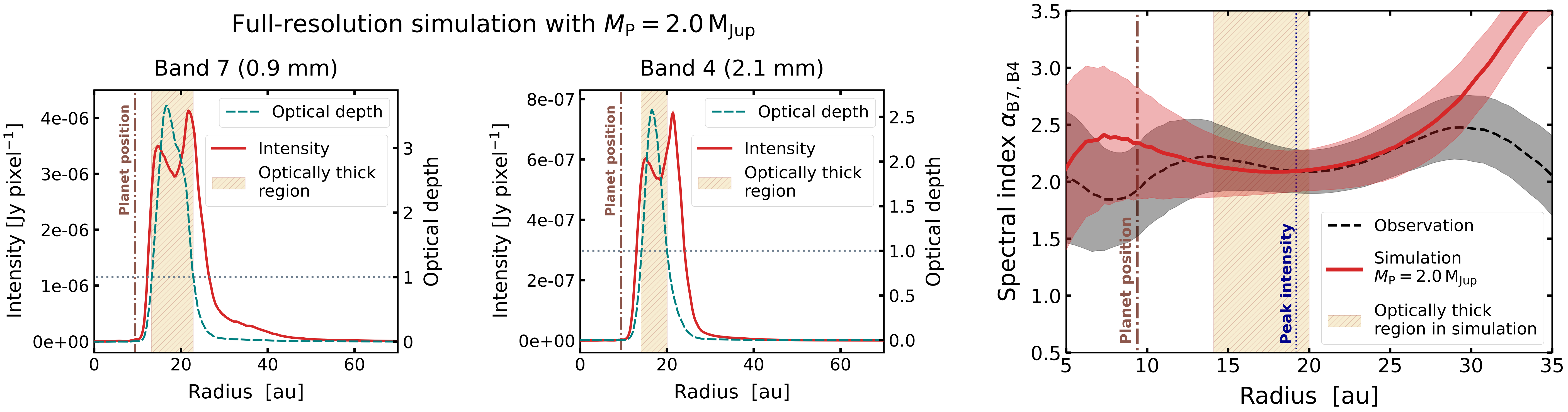}
      \caption{Optical depth in our models and comparison between observed and simulated spectral indices. \textit{Left and central panels}: Azimuthally averaged radial profiles of the optical depth along the line of sight and the intensities from the full-resolution dust emission image of the simulation  (i.e., not convoluted with the synthesized beam)  with a final planet mass of $2.0 \; \mathrm{M}_{\mathrm{Jup}}$ after $\SI{4e4}{\mathrm{yrs}}$, in Band~7 and Band~4. The vertical dash-dotted  brown line indicates the location of the planet, and the horizontal dotted gray line marks the optical depth $\tau = 1$ level. The region where the emission is optically thick (optical depth $\tau \geq 1$) is shown in beige. \textit{Right panel}:  Azimuthally averaged radial profiles of the spectral index from the ALMA observation and our simulation with a final planet mass of $2.0 \; \mathrm{M}_{\mathrm{Jup}}$ after $\SI{4e4}{\mathrm{yrs}}$. Values are obtained using Eq.~\ref{eq:spectral_index} and the synthetic image intensity profiles in Fig.~\ref{fig:mod&res_profiles}. Uncertainties include the original error in the intensity profiles and a 10\% flux error due to calibration. As a reference, the distance of the planet from the central star is indicated by the vertical  dash-dotted  brown line, and the position of the peak intensity is displayed by the vertical dotted blue line. The beige-shaded area marks the radii where the full-resolution simulated emission is optically thick ($\tau \geq 1$) in both Band~7 and Band~4.}
         \label{fig:opticaldepth_spectalindex}
   \end{figure*}

Having dust emission data at two different wavelengths allows us to compare the observed and simulated spectral index. Given the fluxes and frequencies corresponding to Band~7 and Band~4, respectively $F_{\mathrm{B7}}$, $\nu_{\mathrm{B7}}$, and $F_{\mathrm{B4}}$, $\nu_{\mathrm{B4}}$, the spectral index $\alpha_{\mathrm{B7},\,\mathrm{B4}}$ is calculated as
\begin{equation}
    \alpha_{\mathrm{B7},\,\mathrm{B4}} = \frac{\log_{10}\left[F_\mathrm{B7} / F_\mathrm{B4} \right]}{{\log_{10}\left[\nu_\mathrm{B7} / \nu_\mathrm{B4}\right]}}\,.
    \label{eq:spectral_index}
\end{equation}
Assuming that opacity follows a power-law $\kappa_\nu\propto \nu^{\beta}$, the emission is optically thin and within the Rayleigh-Jeans regime, then the spectral index can be approximated as $\alpha_{\mathrm{B7},\,\mathrm{B4}} \approx 2+\beta$ \citep{2014prpl.conf..339T}.  
The value of $\beta$ depends on the maximum dust grain size, while not being influenced by the minimum grain size. A low $\beta$ (i.e., $\beta<1$, meaning that $\alpha_{\mathrm{B7},\,\mathrm{B4}} < 3$) could be indicative of the presence of grains  with size $\gtrsim \SI{0.1}{\mm}$, suggesting that grain growth in the disk may have occurred  \citep{2006ApJ...636.1114D}. Protoplanetary disks around VLM~stars or brown dwarfs with low spectral indices have been observed (e.g., \citealt{2014ApJ...791...20R, 2017ApJ...846...70P}). 

The left and central panels of Fig.~\ref{fig:opticaldepth_spectalindex} present the azimuthally averaged radial profiles of the intensity from our simulations with final planet mass of $2.0 \; \mathrm{M}_{\mathrm{Jup}}$ in Band~7 and Band~4, along with the corresponding profiles of the optical depth along the line of sight. All profiles were produced directly from the full-resolution images computed by \textsc{Polaris}, without any convolution with a synthesized beam. Here, we note that the emission at both wavelengths is partially optically thick, that is, the optical depth $\tau$ is above 1. We calculate that ${\approx}51\%$ of the emission in Band~7 is optically thick, whereas this fraction decreases to ${\approx}38\%$ in Band~4. These results are consistent with the findings in \citet{2021MNRAS.506.2804T}, showing similar optically thick fractions in a sample of disks located in the Lupus star-forming region. Such a mixture of optically thin and thick emission in our model leads us to interpret with caution the spectral index in CIDA~1. Moreover, the dust temperature predicted in our simulation (${\sim}\SI{20}{\K}$ in the disk midplane, where most of the settled dust is located) lies at the limit of the Rayleigh-Jeans domain. In this spectral region, and for a temperature of $\SI{20}{\K}$, dropping the Rayleigh-Jeans approximation but retaining the optically thin one, we calculate that the spectral index is  $\alpha_{\mathrm{B7},\,\mathrm{B4}} \approx 1.7+\beta$.

   \begin{figure*}[]
\centering
\includegraphics[width=0.85\textwidth]{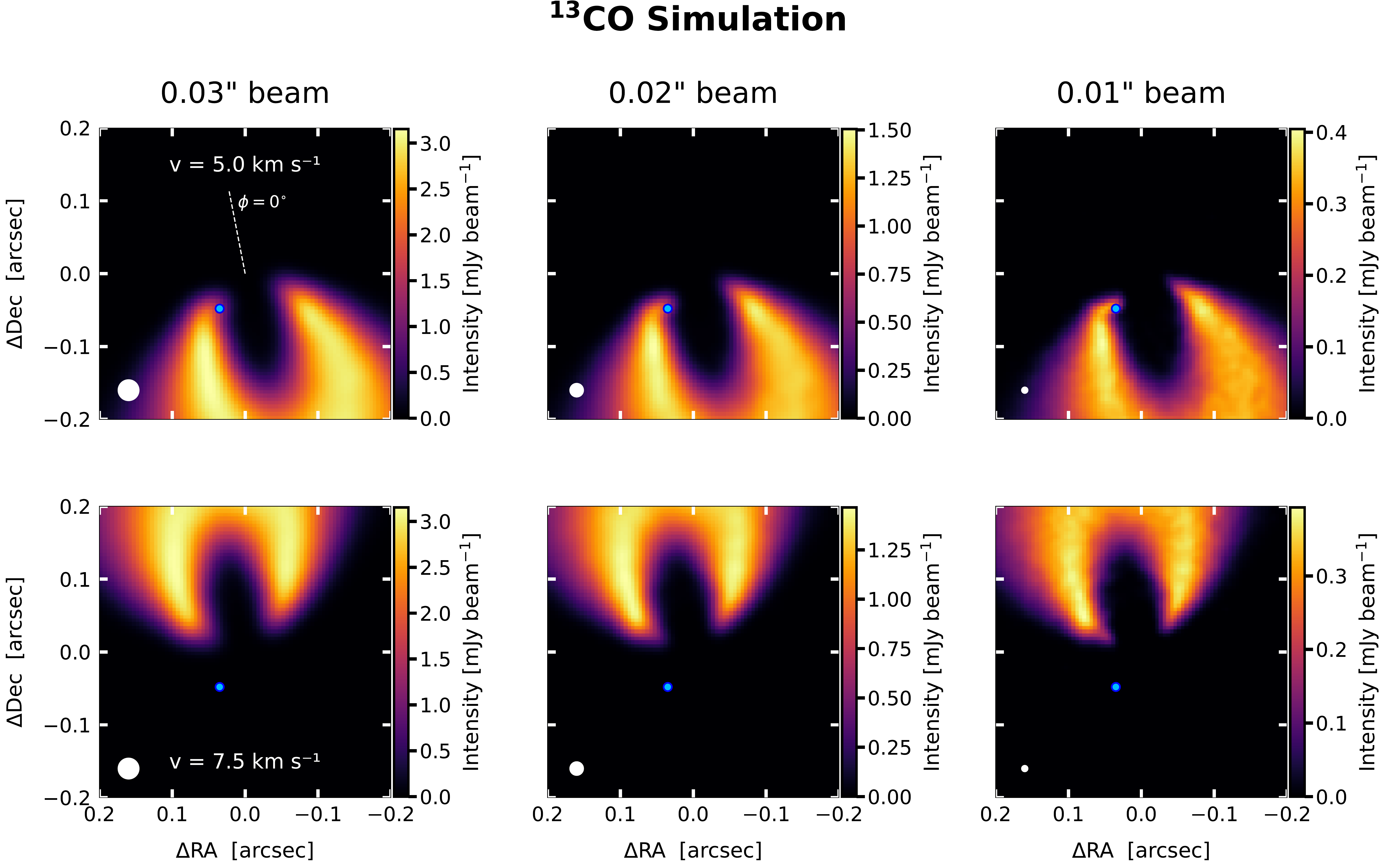}
  \caption{Synthetic images of the $^{13}$CO ($J=3{-}2$) emission from our simulation with $M_{\mathrm{P}}=2.0\,\mathrm{M}_{\mathrm{Jup}}$ after an evolution time of $\SI{4e4}{\mathrm{yrs}}$. We compute the images at two different channels: $\varv=\SI{5.0}{\km\per\s}$ (\textit{top panels}) and $\varv=\SI{7.5}{\km\per\s}$ (\textit{bottom panels}), with a velocity resolution of $\SI{0.5}{\km\per\s}$. We use three different circular beam sizes: $0.03\arcsec$, $0.02\arcsec$, and $0.01\arcsec$ (FWHMs, \textit{from left to right}), indicated by the white circles in the lower-left corner of each plot. The blue dot represents the planet position at the azimuthal angle $\phi=145^{\circ}$ (relative to the angle $\phi=0^{\circ}$, which is located at the disk major axis indicated by the dashed white line in the \textit{top-left panel}).}
  \label{fig:search_kink}
\end{figure*}

\citet{2021A&A...649A.122P} find that the spatially integrated spectral index of CIDA~1 is $2.0\pm0.2$ (the uncertainty includes the observed root mean square of the noise and a 10\% flux error from calibration). Table~\ref{table:spectral_index_spatially_integrated} compares the flux in Band~7, Band~4, and the spatially integrated spectral index between the observation (considering only flux coming from the external dust ring) and our simulation with a final planet mass of $2.0\,\mathrm{M}_{\mathrm{Jup}}$ after $\SI{4e4}{\mathrm{yrs}}$. We note that the spatially integrated spectral index of our model is consistent with the observed value within the uncertainties. 

A more detailed comparison is presented in the right panel of Fig.~\ref{fig:opticaldepth_spectalindex}, which shows the azimuthally averaged radial profiles of the spectral index from the observation and, again, from the simulation with a final planet mass of $2.0\,\mathrm{M}_{\mathrm{Jup}}$ after $\SI{4e4}{\mathrm{yrs}}$. Our model best reproduces the observed spectral index profile between ${\sim}\SI{10}{\mathrm{au}}$ and ${\sim}\SI{30}{\mathrm{au}}$, where the dust ring is detected. Since we focus on reproducing the external dust ring, we do not display the spectral index profiles within a distance of $\SI{5}{\mathrm{au}}$ from the central star. Then, we note that the spectral indices start diverging beyond ${\sim}\SI{30}{\mathrm{au}}$. This effect is due to a slight intensity excess in the model in Band~7 that does not perfectly match the observed Band~7 profile at these radii (see Fig.~\ref{fig:mod&res_profiles}). Between ${\sim}\SI{14}{\mathrm{au}}$ and ${\sim}\SI{20}{\mathrm{au}}$ the dust emission is optically thick in both Band~7 and Band~4, meaning that the low spectral index in this region cannot be linked to the presence of millimeter-sized grains. However, this is not the case for the region between ${\sim}\SI{20}{\mathrm{au}}$ and ${\sim}\SI{30}{\mathrm{au}}$, still showing a spectral index around ${\sim}2{-}2.5$ associated with optically thin emission. Such behavior is expected in our simulation, where we know that millimeter-sized grains are present at the dust ring location, but this strong match between our model and the observation can indicate that grain growth and migration have occurred in CIDA~1.

\subsection{Planet-induced perturbations in synthetic gas channel maps}

The presence of substructures, such as rings, gaps, and cavities, in the dust continuum images of protoplanetary disks can be interpreted as the result of tidal interactions between the disk and a planet. A more direct way to reveal the effects of planet-disk interactions is to look for localized velocity perturbations in the gas channel maps, namely ‘‘kinks" \citep{2018ApJ...860L..13P, 2019NatAs...3.1109P, 2020ApJ...890L...9P,2021A&A...650A.179I, 2021MNRAS.504.5444B}. 

In the channel maps of CIDA~1 observed by ALMA in $^{12}$CO ($J=3{-}2$) and $^{13}$CO ($J=3{-}2$)  (top panels in Fig.~\ref{fig:12CO_channels_comparison} and Fig.~\ref{fig:13CO_channels_comparison}, respectively), there is no clear sign of a deviation from the Keplerian rotation. Here, we investigate whether, with  sufficient sensitivity and both spectral and spatial resolution, a perturbation could be detected in one of the channels that are not affected by cloud absorption.

      \begin{figure*}[]
   \centering
   \includegraphics[width=0.72\textwidth]{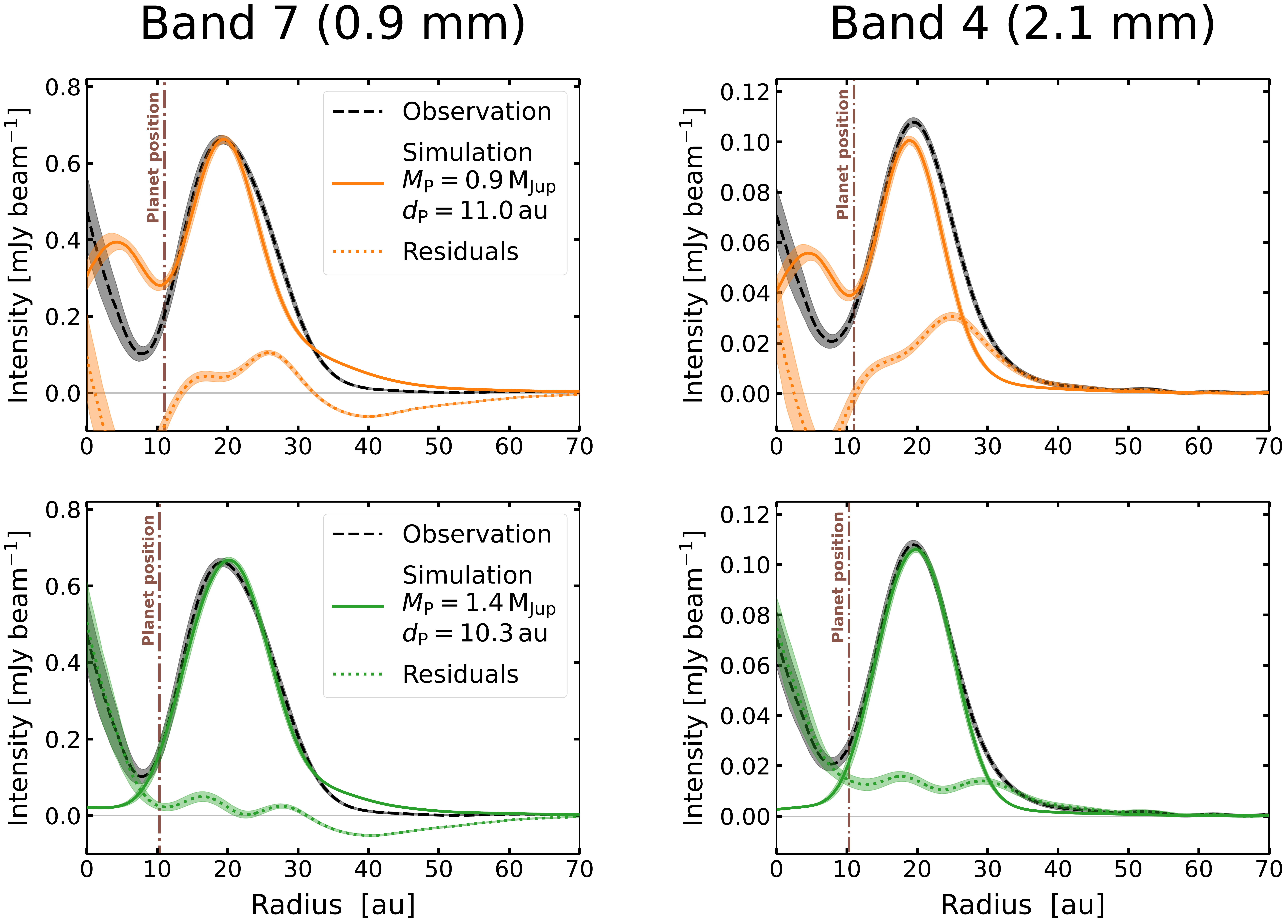}
      \caption{As in Fig.~\ref{fig:mod&res_profiles}, but for the cases of simulations with final planet masses $0.9$ and $1.4\,\mathrm{M}_{\mathrm{Jup}}$ with different initial distances from the star. For the simulation in the \textit{first row},  $M_{\mathrm{P},0} = 0.5\,\mathrm{M}_{\mathrm{Jup}}$ and  $d_{\mathrm{P},0} = \SI{12.0}{\mathrm{au}}$, and, after an evolution time of $\SI{4e4}{\mathrm{yrs}}$, $M_{\mathrm{P}} = 0.9\,\mathrm{M}_{\mathrm{Jup}}$ and  $d_{\mathrm{P}} = \SI{11.0}{\mathrm{au}}$. \textit{Second row}: $M_{\mathrm{P},0} = 1.0\,\mathrm{M}_{\mathrm{Jup}}$ and  $d_{\mathrm{P},0} = \SI{11.0}{\mathrm{au}}$, and, after the same evolution time, $M_{\mathrm{P}} = 1.4\,\mathrm{M}_{\mathrm{Jup}}$ and  $d_{\mathrm{P}} = \SI{10.3}{\mathrm{au}}$.}
         \label{fig:mod&res_profiles_smallest_mass}
   \end{figure*}

We consider our simulation with a final planet mass of $2.0\,\mathrm{M}_{\mathrm{Jup}}$ after $\SI{4e4}{\mathrm{yrs}}$. We explore different velocity resolutions and beam sizes. The synthetic observations are computed by convolving the full-resolution images obtained from the radiative transfer simulations with a circular Gaussian beam.  In the $^{12}$CO emission, there is no evidence of kinks in any channels.  On the other hand, a local perturbation nearby the planet is visible in the $^{13}$CO emission. Figure~\ref{fig:search_kink} compares the simulated $^{13}$CO emission at three different angular resolutions: $0.03\arcsec$, $0.02\arcsec$, and $0.01\arcsec$. We show the channels at $\varv=\SI{5.0}{\km\per\s}$ and $\varv=\SI{7.5}{\km\per\s}$. These velocities, equidistant from the systemic velocity, are just outside the window where the cloud totally absorbs $^{13}$CO emission. We only show the results with a channel width of $\SI{0.5}{\km\per\s}$ since  a higher velocity resolution does not lead to any major change. The planet is located at the azimuthal angle $\phi=145^{\circ}$ with respect to the disk major axis (see Fig.~\ref{fig:search_kink}). We note that a velocity perturbation near the planet is visible at a resolution of $0.02\arcsec$ in the $\SI{5.0}{\km\per\s}$ channel, becoming more evident at $0.01\arcsec$. Instead, no kink is detected in the $\SI{7.5}{\km\per\s}$ channel. Therefore, this analysis proves that a planet-induced kink would be visible only in $^{13}$CO emission and at a very high resolution of at least $0.02\arcsec$, assuming that the planet is located in a favorable position that allows the perturbation to be detected in a channel not obscured by the cloud. The sensitivity needed to clearly detect  $^{13}$CO emission at such a high angular resolution leads to an inaccessible observing time (about several tens of days) with current ALMA capabilities. 

We identify two main reasons why the planet-induced perturbation is only detected in $^{13}$CO and not in $^{12}$CO in our simulations. First, the emission of $^{12}$CO comes from regions at high altitudes, whose dynamics could be less affected by the planet orbiting on the midplane \citep{2021MNRAS.502.5325R}, whereas $^{13}$CO is emitted closer to the midplane.  Second, the spatial resolution of our models naturally becomes coarser at higher altitudes, where $^{12}$CO is emitted, thus making it more challenging to detect localized perturbations induced by a planet. So far, published detections of planet-induced kinematical kinks in real ALMA observations of disks have involved only $^{12}$CO \citep{2018ApJ...860L..13P, 2019NatAs...3.1109P, 2020ApJ...890L...9P}. The explanation is that $^{12}$CO has a bright line emission, allowing for very high signal-to-noise ratios even at high spectral and spatial resolution. Kinks in $^{13}$CO have a larger kinematical signal, but are more difficult to detect because of its fainter line brightness.   

\subsection{Minimum planet mass}
\label{sect:min_MP}

The comparison in Fig.~{\ref{fig:mod&res_profiles}} shows that  final planet masses of  $0.9 \; \mathrm{M}_{\mathrm{Jup}}$ and $1.4 \; \mathrm{M}_{\mathrm{Jup}}$ are able to open a gap in the dust, though producing a ring that is shifted with respect to the observed one. While all planets considered in Fig.~{\ref{fig:mod&res_profiles}} have an equal initial distance $d_{\mathrm{P},0}$ from the central star of $\SI{10.0}{\mathrm{au}}$, now, we aim to better constrain the minimum planet mass able to explain the observed dust ring by modifying the parameter $d_{\mathrm{P},0}$. Using the same initial planet masses that eventually resulted in $0.9 \; \mathrm{M}_{\mathrm{Jup}}$ and $1.4 \; \mathrm{M}_{\mathrm{Jup}}$, that is, $M_{\mathrm{P},0} = 0.5$  and $1.0 \; \mathrm{M}_{\mathrm{Jup}}$, respectively, we explore different values of $d_{\mathrm{P},0}$. The best cases are presented in Fig.~{\ref{fig:mod&res_profiles_smallest_mass}}, after an evolution time of $\SI{4e4}{\mathrm{yrs}}$. 

The planet with $M_{\mathrm{P}} = 0.9\,\mathrm{M}_{\mathrm{Jup}}$ and  $d_{\mathrm{P}} = \SI{11.0}{\mathrm{au}}$ is not able to reproduce the observed profiles. As explained in Sect.~\ref{sect:focus_external_disk}, our numerical resolution prevents us from trusting the dust morphology inside the first few astronomical units of the disk. In this case, therefore, we do not deem the apparent presence of an inner ring within 4 au significant (which might be reminiscent of the observed inner disk in CIDA 1), but we clearly see an excess emission between ${\sim} 4$ and ${\sim} \SI{10}{\mathrm{au}}$ (a region properly mapped by SPH particles in our simulation), which is significantly higher than what is observed. In addition, the morphology of the external dust ring is slightly different, particularly in Band~4. Conversely, we note an excellent match between the observed dust ring and the one obtained from a planet with $M_{\mathrm{P}} = 1.4\,\mathrm{M}_{\mathrm{Jup}}$ and  $d_{\mathrm{P}} = \SI{10.3}{\mathrm{au}}$ in both Band~7 and Band~4. Therefore, we conclude that a minimum mass of ${\sim}1.4\,\mathrm{M}_{\mathrm{Jup}}$ can explain the observed dust ring.

Such constraints on the planet mass have been obtained by varying the initial planet mass and location but always fixing the initial value of the disk viscosity at $\alpha_{\mathrm{SS}} = 5\times10^{-3}$. Through a set of hydrodynamical simulations of disks with a planet, \citet{2018A&A...612A.104F} prove that the dust gap width (which they define as the distance between the peak of the dust ring and the center of the gas gap) remains substantially unchanged when varying the disk viscosity by an order of magnitude (from  $\alpha_{\mathrm{SS}} = 10^{-3}$ to $10^{-4}$), thus showing its effective independence from viscosity (see Fig.~11 in their paper). On the contrary, the CO gap width appears to be dependent on viscosity (Fig.~13 in \citealt{2018A&A...612A.104F}).

Beyond viscosity, in our simulations we maintained the same initial values of all the other parameters characterizing the disk (see Table~\ref{table:Phantom_params}) and the same dust composition. Given the computational cost of each simulation, a full exploration of all these parameters is out of the scope of this work.

\subsection{Maximum planet mass}
\label{sect:max_MP}

   \begin{figure*}[]
   \centering
   \includegraphics[width=0.72\textwidth]{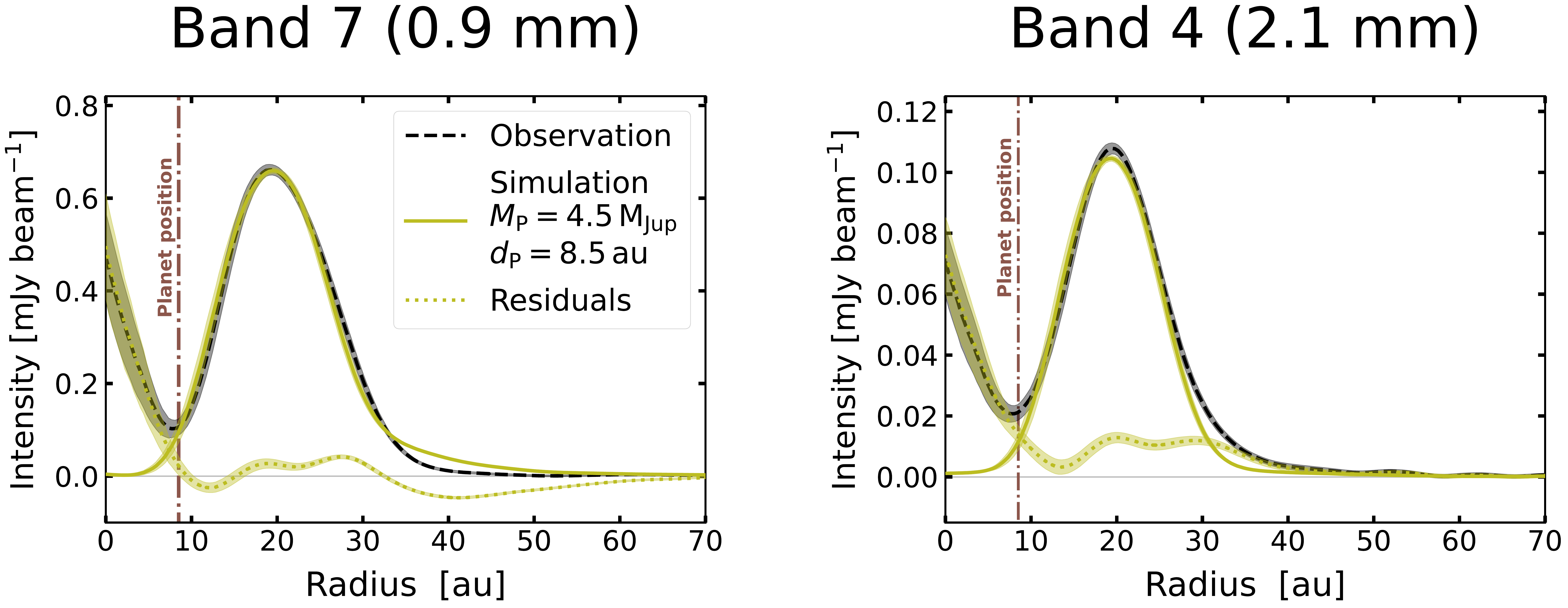}
      \caption{As in Fig.~\ref{fig:mod&res_profiles}, but for the case of a simulation with higher planet mass. The planet started with $M_{\mathrm{P},0} = 4.0\,\mathrm{M}_{\mathrm{Jup}}$ and  $d_{\mathrm{P},0} = \SI{9.0}{\mathrm{au}}$, and, after an evolution time of $\SI{4e4}{\mathrm{yrs}}$, it reached $M_{\mathrm{P}} = 4.5\,\mathrm{M}_{\mathrm{Jup}}$ and  $d_{\mathrm{P}} = \SI{8.5}{\mathrm{au}}$.}
         \label{fig:mod&res_profiles_4MJup}
   \end{figure*}  

   \begin{figure}[]
   \centering
   \includegraphics[width=\hsize]{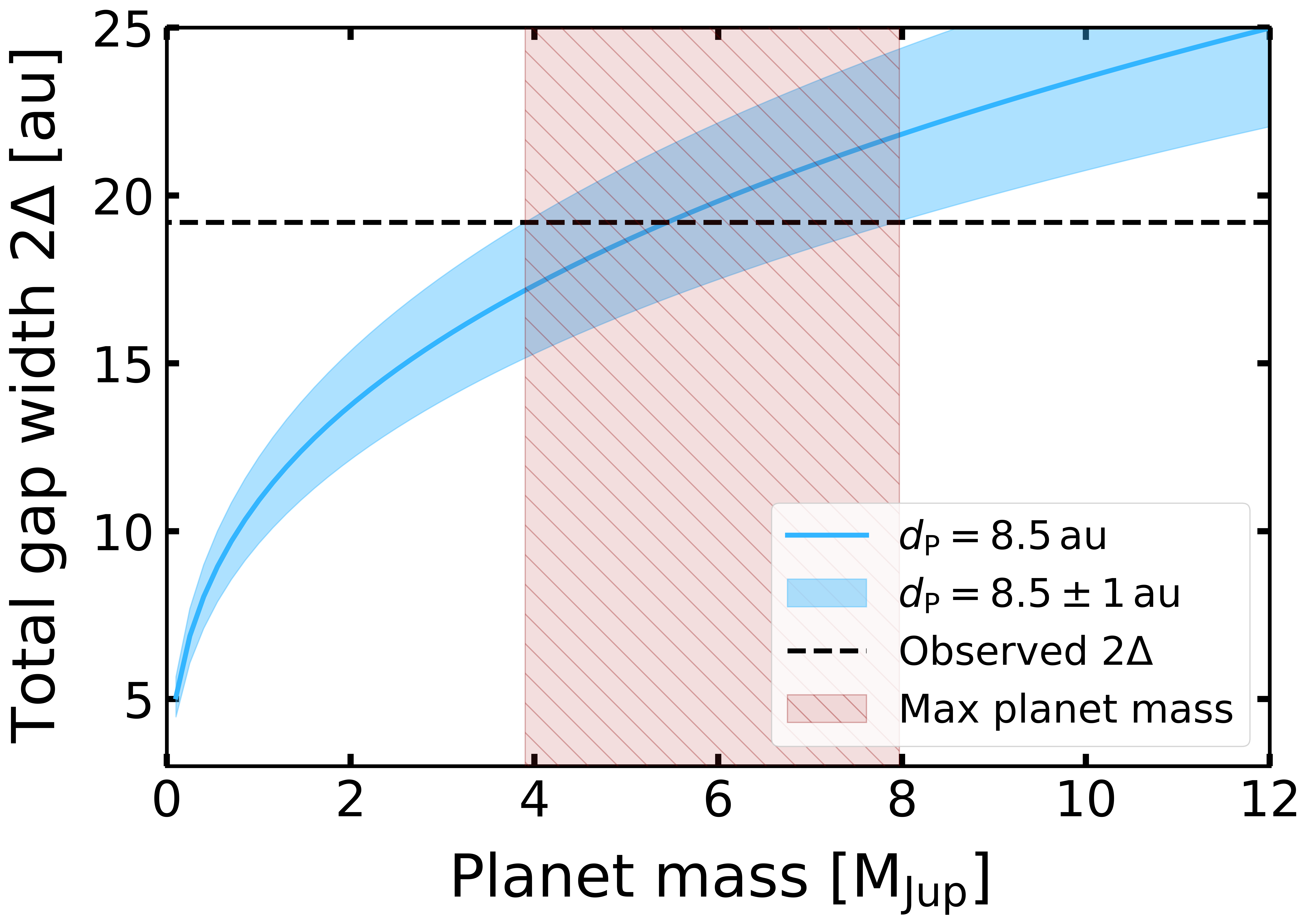}
      \caption{Total gap width in the dust as a function of the planet mass, derived from Eq.~\eqref{eq:Lodato2019}. The red-shaded area indicates the maximum planet mass that could fit the observed total gap width, assuming that the planet is located between $7.5$ and $\SI{9.5}{\mathrm{au}}$ from the star.. 
              }
         \label{fig:max_planet_mass_Lodato2019}
   \end{figure}
   
The simulations reported in Fig.~\ref{fig:mod&res_profiles} do not allow an estimation of the maximum planet mass able to recover the dust ring in the observations. The cases with the highest final planet masses, $2.0$ and $2.5\,\mathrm{M}_{\mathrm{Jup}}$,  starting at a distance of $\SI{10.0}{\mathrm{au}}$ from the central star, both produce an accurate match with the observed dust emission morphology. We try to constrain the maximum planet mass by simulating a new case with a higher initial planet mass of  $M_{\mathrm{P},0} = 4.0\,\mathrm{M}_{\mathrm{Jup}}$. Choosing an initial location $d_{\mathrm{P},0} = \SI{9.0}{\mathrm{au}}$, after $\SI{4e4}{\mathrm{yrs}}$ the planet reached  $d_{\mathrm{P}} = \SI{8.5}{\mathrm{au}}$ with $M_{\mathrm{P}} = 4.5\,\mathrm{M}_{\mathrm{Jup}}$, and also this model reproduces the observed dust ring in both Band~7 and Band~4 (see Fig.~\ref{fig:mod&res_profiles_4MJup}). 

Thus, we followed another approach to give an estimate of the possible maximum planet mass. \citet{2019MNRAS.486..453L} report an empirical relation according to which the width of an observed dust gap scales  with the planet Hill radius (see Eq.~\ref{eq:Hill_radius}) in case of low disk viscosity ($\alpha_{\mathrm{SS}}\lesssim 0.01$; \citealt{2011ApJ...738..131D, 2012A&A...545A..81P, 2016MNRAS.459.2790R, 2016ApJ...832..105F, 2018A&A...612A.104F}). Averaging results from hydrodynamical simulations \citep{2018ApJ...866L...6C, 2019A&A...622A..75L}, they derive
\begin{equation}
    \Delta = 5.5 R_{\mathrm{H}} \, ,
    \label{eq:Lodato2019}
\end{equation}
where $\Delta$ is defined as the distance between the minimum intensity in the gap and the peak intensity in the ring. Hence, we take the value of $2\Delta$ as the total dust gap width. This relation assumes that the planet position coincides with the location of the dust gap, and that the gap is opened by a single planet. With such an
analytical approach, we do not suffer from the numerical
limitations of the hydrodynamical simulations (see Sect.~\ref{sect:focus_external_disk}). Instead,  we can
place an upper limit to the planet mass by requiring the dust gap that it produces not to be so wide as to prevent the formation of the observed inner disk (which we did not reproduce in the hydrodynamical simulations).

From the intensity radial profile of CIDA~1 observed in both Band~7 and Band~4 (see Fig.~\ref{fig:mod&res_profiles}), the total dust gap width is $\SI{19.2}{\mathrm{au}}$. We measured this distance from the peak of the inner disk emission (i.e., the center of the disk) to the peak of the dust ring. We assumed that the observed inner disk is not a fast transient feature, but a long-lived structure. We fixed the star mass at $\SI{0.2}{M_{\odot}}$ and adopted a planet location between $\SI{9.5}{\mathrm{au}}$, consistent with the values in Table~\ref{tab:evolution_SPH_params}, and  $\SI{7.5}{\mathrm{au}}$, assuming that a higher-mass planet needs a shorter distance from the star to correctly reproduce the observed dust ring, as occurred with the case of the $4.5\,\mathrm{M}_{\mathrm{Jup}}$ planet (Fig.~\ref{fig:mod&res_profiles_4MJup}). Employing Eq.~\ref{eq:Lodato2019} with these values, we obtained the total dust gap width $2\Delta$ versus the planet mass, as depicted in Fig.~\ref{fig:max_planet_mass_Lodato2019}. Comparing the computed gap widths with the observed one, we find that the maximum planet mass able to create the CIDA~1 gap should be ${\sim}4{-}8\,\mathrm{M}_{\mathrm{Jup}}$. A more massive planet would carve a wider gap, either  generating the dust ring at a greater distance from the star or depleting the inner disk.

\subsection{Comparison with previous estimates of the planet mass}

Assuming a stellar mass of $0.1\,\mathrm{M}_\odot$, a disk viscosity $\alpha_{\mathrm{SS}}=10^{-3}$, and that the gap is carved by a single planet located at the minimum of the gap (${\sim}\SI{8}{\mathrm{au}}$), \citet{2021A&A...649A.122P} employed the \citet{2006Icar..181..587C} criterion and estimated that the minimum planet mass to open a gap in the gas is ${\sim} 0.4\,\mathrm{M_{\mathrm{Jup}}}$. However, such a planet would generate a pressure maximum at ${\sim}\SI{13}{\mathrm{au}}$. So, either a more massive planet or a higher distance from the star is needed to retrieve the observed dust ring at ${\sim}\SI{20}{\mathrm{au}}$. Following this argument, the authors computed dust evolution simulations considering two cases: a $2.4\,\mathrm{M}_{\mathrm{Jup}}$ planet located at $\SI{8}{\mathrm{au}}$ and a $0.5\,\mathrm{M}_{\mathrm{Jup}}$ planet at $\SI{12}{\mathrm{au}}$, assuming a gap shape obtained analytically. Their 1D models evolved for $\SI{e6}{\mathrm{yrs}}$, starting from  micron-sized dust particles and then considering dust growth, fragmentation and erosion. Both planets were able to form a dust ring at the observed location. The $2.4\,\mathrm{M}_{\mathrm{Jup}}$ planet case was excluded: the formed gap was too deep to allow replenishment of dust from the outer region to an inner disk, resulting in an empty internal cavity. The $0.5\,\mathrm{M}_{\mathrm{Jup}}$ planet case, instead, was able to reproduce the contrast between the inner disk and the ring after $\SI{e5}{\mathrm{yrs}}$ of evolution. At longer times, however, the inner disk became fainter.

The models presented in this work differ significantly from those in \citet{2021A&A...649A.122P}. Our hydrodynamical simulations take into account all the complex dynamical interactions between the disk gas and dust components and the planet in 3D. The computational effort needed inevitably leads to evolution times shorter than $\SI{e6}{\mathrm{yrs}}$. We did not include the effects of dust evolution such as grain growth or fragmentation, but we assumed that dust growth has already occurred, simulating the dynamical behavior of different dust populations with fixed grain sizes, ranging from submicron to millimeter scales. In our models, most of the dust mass is contained in the bigger grain sizes, allowing us to reproduce the observed low value of the spectral index. On the other hand, the simulations of \citet{2021A&A...649A.122P} struggled to produce large dust grains, leading to  higher values of the spatially-integrated spectral index (${\sim}2.6$). The numerical limitations described in Sect.~\ref{sect:focus_external_disk} prevent us from reproducing the inner disk, but the final synthetic images in our models allow, nonetheless, a careful match with the observed dust ring and the gas emission morphology. Moreover, thanks to the comparison between the simulated and observed gas channel maps, we find that a better estimate of the stellar mass is $0.2\,\mathrm{M}_\odot$, whereas \citet{2021A&A...649A.122P}
assumed a $0.1\,\mathrm{M}_\odot$ central star. This difference in the stellar mass, along with a disk viscosity $\alpha_{\mathrm{SS}}=10^{-3}$, lower than the initial value of $5\times10^{-3}$ adopted in our simulations, might explain why \citet{2021A&A...649A.122P} predict a planet mass of ${\sim}0.5\,\mathrm{M}_{\mathrm{Jup}}$ while our models require a minimum planet mass of ${\sim}1.4\,\mathrm{M}_{\mathrm{Jup}}$ to reproduce the observed dust ring.

The two modeling approaches are different, and neither is fully complete. Future studies  need to take into account the phenomena of dust growth and fragmentation within  comprehensive hydrodynamical and radiative transfer simulations. Including all of these effects, each with its typical timescale, may end up reducing the uncertainty on the mass of the planet that can generate the observed substructures in CIDA~1. Moreover, it is important to note that our work and the one by \citet{2021A&A...649A.122P} aim at reproducing the observed disk morphology with a single planet. It remains to be explored whether a multi-planet system could reproduce the disk emission, and future simulations including more than one embedded planet are needed to test such a scenario.

\subsection{Implications for planet formation around VLM stars}

Our analysis of the CIDA~1 system is consistent with the presence of a planet more massive than Jupiter around a $0.2\,\mathrm{M}_\odot$ star. Even considering that CIDA~1 may be an outlier in terms of the initial disk mass or its properties, it can  still provide relevant constraints on planet formation theories. It is thus important to compare what we find in this system with the prediction of the leading theories.

Core accretion model \citep{1996Icar..124...62P} is based on the dynamics of dust in disks and how it evolves from submicron dust grains in the interstellar medium to kilometer-sized planetesimals through collisions. During this process, when dust grains reach millimeter size, they should rapidly drift toward the central star, preventing grains from growing into planetesimals and planetary cores \citep{1977MNRAS.180...57W}. This barrier for planet formation is even harder to overcome for disks around VLM stars since dust radial drift is more efficient in these environments \citep{2013A&A...554A..95P, 2018MNRAS.479.1850Z}. Furthermore, the disk mass content is crucial to determine whether planet formation may occur. Disk population studies in nearby star-forming regions (e.g.,  \citealt{2016ApJ...827..142B,2016ApJ...828...46A, 2016A&A...593A.111T, 2016ApJ...831..125P,2019MNRAS.482..698C, 2019ApJ...875L...9W,  2020A&A...633A.114S,2021A&A...653A..46V}) prove that the scaling relation between disk mass and stellar mass is steeper than linear. Therefore, it appears that disks in the low-stellar-mass regime do not possess enough material to form the known planetary systems \citep{2018A&A...618L...3M}, unless the planet-formation efficiency is 100\% \citep{2021ApJ...920...66M}. However, these studies consider disks with a mean age $> \SI{e6}{\mathrm{yrs}}$, whereas planet formation via gravitational instabilities should occur at earlier times \citep{2016ARA&A..54..271K}.
The properties of disks in the early Class~0/I protostars is  an active area of research and the properties of such disks are highly uncertain and debated at the moment, both on observational and modeling grounds \citep{2019A&A...621A..76M,2020A&A...640A..19T,2021ApJ...917L..10L}. Nevertheless, it seems likely that at least some young disks will be prone to undergo gravitational instabilities and possibly form massive planets early on through this path.

Various theoretical studies employing numerical simulations have been performed to investigate planet formation through core accretion or gravitational instability in the low-stellar-mass regime. \citet{2007MNRAS.381.1597P} assessed planet formation around brown dwarfs adapting models for higher stellar masses based on core accretion. Through Monte Carlo simulations, they found that Earth-like planets can form in this condition and the planet mass depends strongly on the disk mass. However, none of their simulations showed a planetary rocky core accreting a gaseous envelope to form a giant planet. A way to overcome the radial drift barrier is a rapid rocky core growth. Pebble accretion is a mechanism able to speed up significantly the giant planet formation process (i.e., \citealt{2012A&A...544A..32L, 2015A&A...582A.112B}). \citet{2020A&A...638A..88L} carried out a theoretical study on planet formation driven by pebble accretion in the (sub)stellar mass range  between $0.01$ and $0.1\,\mathrm{M}_\odot$. First, they calculated the initial masses of protoplanets by extrapolating previous numerical simulations conducted in previous literature. Next, they performed a population synthesis study to track the growth and migration of a large sample of protoplanets under the influence of pebble accretion. Their results show that, around a $0.01\,\mathrm{M}_\odot$ brown dwarf, planets can grow up to $0.1{-}0.2 \, \mathrm{M}_\oplus$, while, around $0.1\,\mathrm{M}_\odot$ stars, planets can reach a maximum mass of $2{-}3\,\mathrm{M}_\oplus$. Findings from this study show that even pebble accretion does not seem to be sufficient to form gas giants around VLM stars and brown dwarfs. \citet{2020MNRAS.491.1998M} used a population synthesis approach based on planetesimal accretion to explore planet formation in the stellar mass range between $0.05$ and $0.25\,\mathrm{M}_{\mathrm{\odot}}$. They let the synthetic population of planetary systems evolve for $\SI{e8}{\mathrm{yrs}}$. The authors find that to form planets with masses higher than $0.1\,\mathrm{M}_\oplus$ they need stars of at least $0.07\,\mathrm{M}_\odot$, implying that planet formation around brown dwarf may not be a usual outcome. Then, stars with masses higher than $0.15\,\mathrm{M}_\odot$ are necessary to form planets more massive than the Earth. Therefore, from all of these studies, we conclude that core accretion model currently cannot explain the presence of gas giants around VLM stars or brown dwarfs. Either the core accretion theory is incomplete, or another mechanism for planet formation is needed. This is confirmed by \citet{2005MNRAS.364L..91L}, who discussed the origin of the $5\,\mathrm{M}_{\mathrm{Jup}}$ planet  detected around the $25\,\mathrm{M}_{\mathrm{Jup}}$ brown dwarf 2MASSW J1207334-393254 \citep{2005A&A...438L..25C}. They found that the core accretion mechanism is far too slow to generate such a planet in less than $\SI{e7}{\mathrm{yrs}}$, the estimated age of the system. Therefore, the authors proposed gravitational instabilities arising  during  the  early  phases  of  the  disk  lifetime as a viable possibility for the formation of the planet.

\citet{2020MNRAS.494.4130H} assessed the susceptibility for disk fragmentation due to self-gravity depending on the properties of the disk and the host star. They performed a set of SPH simulations and accounted for the stellar irradiation. Their results show that disks around VLM stars are less prone to fragment with respect to disks around more massive stars. Fragmentation can occur around VLM stars only if the disk is optically thick to stellar irradiation and there is a high disk-to-star mass ratio (\mbox{$M_{\mathrm{disk}}/M_\star \gtrsim 0.3$}). \citet{2020A&A...633A.116M} focused on the role of disk instability for planet formation around  low-mass stars. They conducted a set of SPH simulations to study the fragmentation conditions of initially gravitationally stable disks, whose mass was slowly but steadily incremented. They considered stars with masses between $0.2$ and $0.4\,\mathrm{M}_\odot$. The authors find that, via gravitational instabilities, protoplanets form very fast, within a few thousand years, provided that the disk is massive enough (\mbox{$M_{\mathrm{disk}}/M_\star \gtrsim 0.3$}). The formed planets  have masses between $2$ and $6\,\mathrm{M}_{\mathrm{Jup}}$ and they are initially distant from the central star, between $15$ and $105\,\mathrm{au}$, but may migrate afterward. Hence, a plausible origin of a giant planet around CIDA~1 might be the fast fragmentation of the disk due to gravitational instabilities, assuming the disk was massive enough in its early stages.

\section{Conclusions}
\label{sect:conclusions}

In this work we inspect the origin of the substructures observed by ALMA in the protoplanetary disk around CIDA~1, one of the very few young VLM stars with high-resolution data. Assuming the presence of an embedded planet, we self-consistently reproduced the observed continuum emission from the dust ring and the gas line emission morphology with a single model. We performed 3D hydrodynamical simulations of gas and multigrain dust, considering initial planet masses from $0.1$ to $4.0\,\mathrm{M}_{\mathrm{Jup}}$ and starting at a radial distance of between $9$ and $11\,\mathrm{au}$ from the central star. Then, we computed the dust and gas emission using radiative transfer modeling. Finally, we treated the obtained images as actual ALMA observations to acquire the final synthetic images. Here we summarize our main findings:
\begin{itemize}
    \item We compared the observed and simulated dust emission profiles in Band~7 (\SI{0.9}{\mm}) and Band~4 (\SI{2.1}{\mm}). We find that the observed dust ring can be explained by the presence of a giant planet  with a minimum mass of ${\sim}1.4\,\mathrm{M}_{\mathrm{Jup}}$ located at a distance of $\SI{10.3}{\mathrm{au}}$ from the central star. The formation of such a massive planet around a VLM star such as CIDA~1 challenges our current theoretical models based on core accretion theory. A valid explanation for its origin may involve gravitational instabilities in the early stages of disk evolution.
    \item Our numerical models cannot directly constrain the maximum planet mass consistent with observations. Thus, we used an empirical relation between the observed dust gap width and the Hill radius of the planet and estimate a maximum planet mass of ${\sim}4{-}8\,\mathrm{M}_{\mathrm{Jup}}$.
    \item Assuming that dust is composed of porous grains with a mixture of astronomical silicates, carbonaceous  materials, and water ice, we match the observed flux with a total dust mass in our models of ${\sim}\SI{7e-6}{\mathrm{M}_\odot}$, corresponding to a dust-to-gas mass ratio of ${\sim}\SI{e-2}{}$. 
    \item The observed $^{12}$CO and $^{13}$CO channel maps are strongly affected by cloud absorption. Nonetheless, our model with a final planet mass of $2.0\,\mathrm{M}_{\mathrm{Jup}}$ can recreate the spatial extent of the gas emission in the channels where it was detected, except for the channels that probe the highest  velocities in the innermost regions, which are not considered in our model. Our simulation include a dust-to-gas mass ratio of ${\sim}{10^{-2}}$, a $^{12}$CO abundance of $[^{12}\mathrm{CO}/\mathrm{H}_2]=5\times10^{-5}$, and an isotopologue abundance ratio $[^{12}\mathrm{CO}/^{13}\mathrm{CO}] = 70$. Since we can recreate the spatial extent and rotation pattern of the disk, we conclude that the assumed stellar mass of $0.2\,\mathrm{M}_{\odot}$ and systemic velocity of $\SI{6.25}{\km\per\s}$ are solid estimates for the actual properties of CIDA~1.
    \item In the case of our simulation with a final planet mass of $2.0\,\mathrm{M}_{\mathrm{Jup}}$, we are able to reproduce the low spectral index (${\sim}2$)  observed at the location of the dust ring ($10{-}30\,\mathrm{au}$). The simulated dust emission in both Band~7 and Band~4 is optically thick in the range $14{-}\SI{20}{\mathrm{au}}$, with a total fraction of optically thick emission corresponding to ${\sim}40{-}50\%$. However, the emission from our model is optically thin between  $20$ and $\SI{30}{\mathrm{au}}$. The low spectral index in this region is expected in our simulations that contain millimeter-sized dust particles, but the match with the observation  hints at the occurrence of grain growth and migration in CIDA~1.
    \item The $2.0\,\mathrm{M}_{\mathrm{Jup}}$ planet creates a velocity perturbation that is predicted to be extremely arduous to observe. It would require $^{13}$CO channel maps at a very high angular resolution of at least $0.02\arcsec$, with a velocity resolution of $\SI{0.5}{\km\per\s}$. 
\end{itemize}

Our results suggest that the observed structures in CIDA~1 can indeed be explained by the presence of a massive embedded planet. It remains to be understood whether CIDA 1 constitutes a unique case or if other disks around VLM stars may also have undergone giant planet formation.

\begin{acknowledgements}
We thank the anonymous referee for the constructive comments that helped improve this paper. This work was partly supported by the Italian Ministero dell Istruzione, Universit\`a e Ricerca through the grant Progetti Premiali 2012 – iALMA (CUP C$52$I$13000140001$), 
by the Deutsche Forschungs-gemeinschaft (DFG, German Research Foundation) - Ref no. 325594231 FOR $2634$/$1$ TE $1024$/$1$-$1$, 
and by the DFG cluster of excellence Origins (www.origins-cluster.de). 
This project has received funding from the European Union's Horizon 2020 research and innovation programme under the Marie Sklodowska-Curie grant agreement No. 823823 (RISE DUSTBUSTERS project) and from the European Research Council (ERC) via the ERC Synergy Grant {\em ECOGAL} (grant 855130).

This paper makes use of the following ALMA data: ADS/JAO.ALMA\#2018.1.00536.S,
ADS/JAO.ALMA\#2015.1.00934.S, and  ADS/JAO.ALMA\#2016.1.01511.S. ALMA is a partnership of ESO (representing its member states), NSF (USA) and NINS (Japan), together with NRC (Canada), MOST and ASIAA (Taiwan), and KASI (Republic of Korea), in cooperation with the Republic of Chile. The Joint ALMA Observatory is operated by ESO, AUI/NRAO and NAOJ.

M.T. has been supported by the UK Science and Technology research Council (STFC) via the consolidated grant ST/S000623/1, and by the European Union’s Horizon 2020 research and innovation programme under the Marie Sklodowska-Curie grant agreement No. 823823 (RISE DUSTBUSTERS project).
\end{acknowledgements}

\bibliographystyle{aa} 
\bibliography{bibliography.bib} 

\begin{appendix}

\section{Sound speed power law exponent from the SED}
\label{appendix:q_parameter}

   \begin{figure}[h]
   \centering
   \includegraphics[width=\hsize]{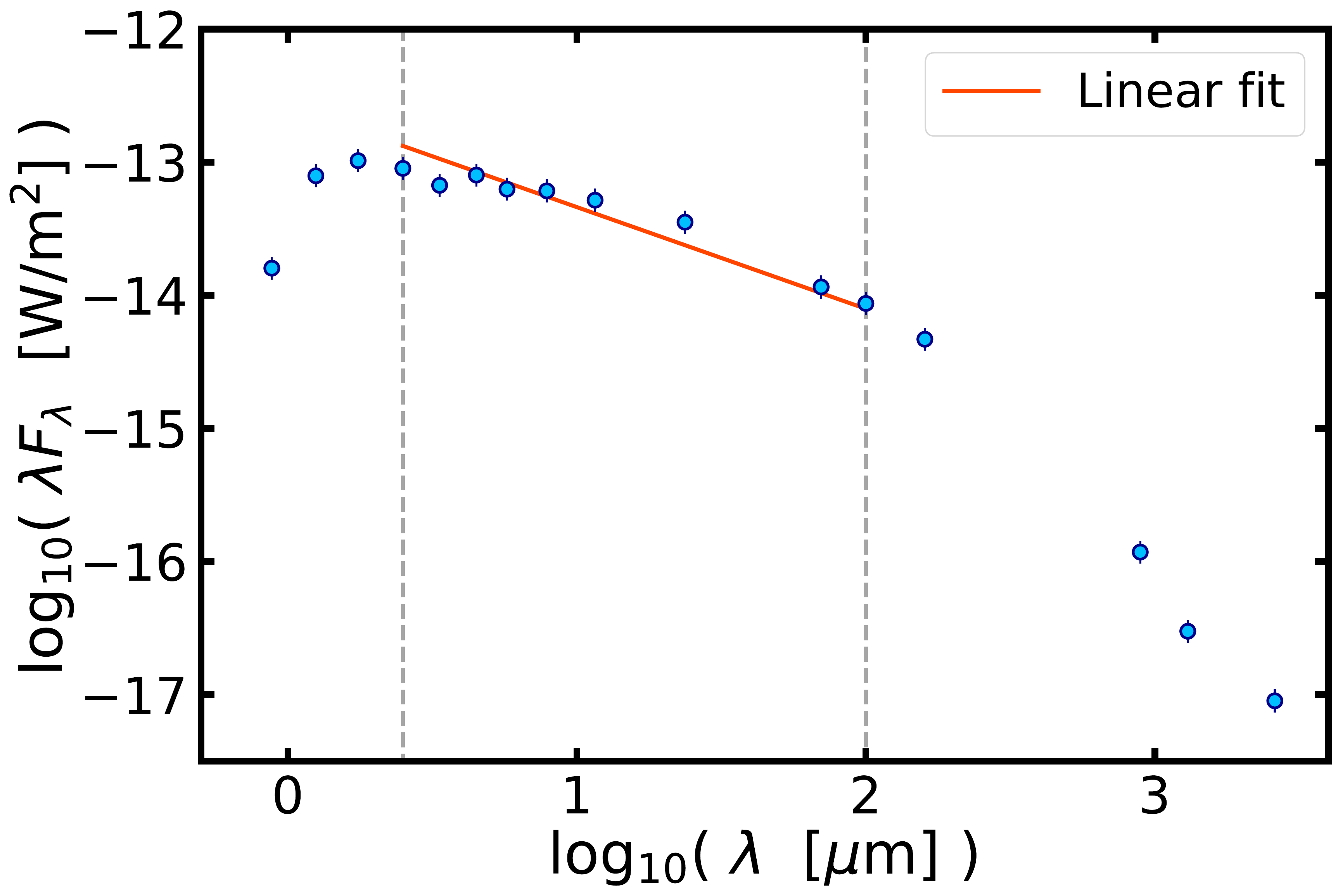}
      \caption{SED of CIDA~1. Data are from \citet{2017ApJ...841..116H}, and  a typical uncertainty of 20\% is assumed for each flux measurement. The orange line represents the linear fit performed on the data in the wavelength range $2.5{-}\SI{100}{\um}$, between the two vertical dashed gray lines.}
         \label{fig:SED}
   \end{figure}
   
The profile of the sound speed in the SPH simulations follows a power law (Eq. \ref{eq:c_s}). It is possible to derive the value of its exponent from the disk spectral energy distribution (SED), assuming that the disk emits as a multicolor blackbody \citep{1990AJ.....99..924B}.  In the frequency range  $k_{\mathrm{B}}T_\mathrm{out} \ll h\nu \ll k_{\mathrm{B}}T_\mathrm{in}$ (where $k_{\mathrm{B}}$ is the Boltzmann constant, $T_\mathrm{in}$ and $T_\mathrm{out}$ are the disk temperatures at the inner and outer radii of the disk, respectively), usually corresponding to a wavelength range roughly between $\SI{2}{\um}$ and $\SI{100}{\um}$, the slope of the SED is connected to the exponent $q$ of the disk sound speed profile by this relation:
\begin{equation}
    \frac{\mathrm{d}\mathrm{\log_{10}}(\lambda\, F_\lambda)}{\mathrm{d}\mathrm{\log_{10}(\lambda)}} = -4+\frac{1}{q}
    \label{eq:SED}
.\end{equation}

The measured SED of CIDA 1 is reported in \citet{2017ApJ...841..116H}. We performed a linear fit of the value of $\mathrm{\log_{10}}(\lambda\, F_\lambda)$ with respect to $\log_{10}(\lambda)$ in the wavelength range $2.5{-}\SI{100}{\um}$ (Fig.~\ref{fig:SED}), obtaining a slope of ${\approx} -0.62$. Using the \eqref{eq:SED}, we derive that $q\approx 0.3$.

\section{Planet accretion and migration}
\label{appendix:planet_evolution}

   \begin{figure}[h]
   \centering
   \includegraphics[width=\hsize]{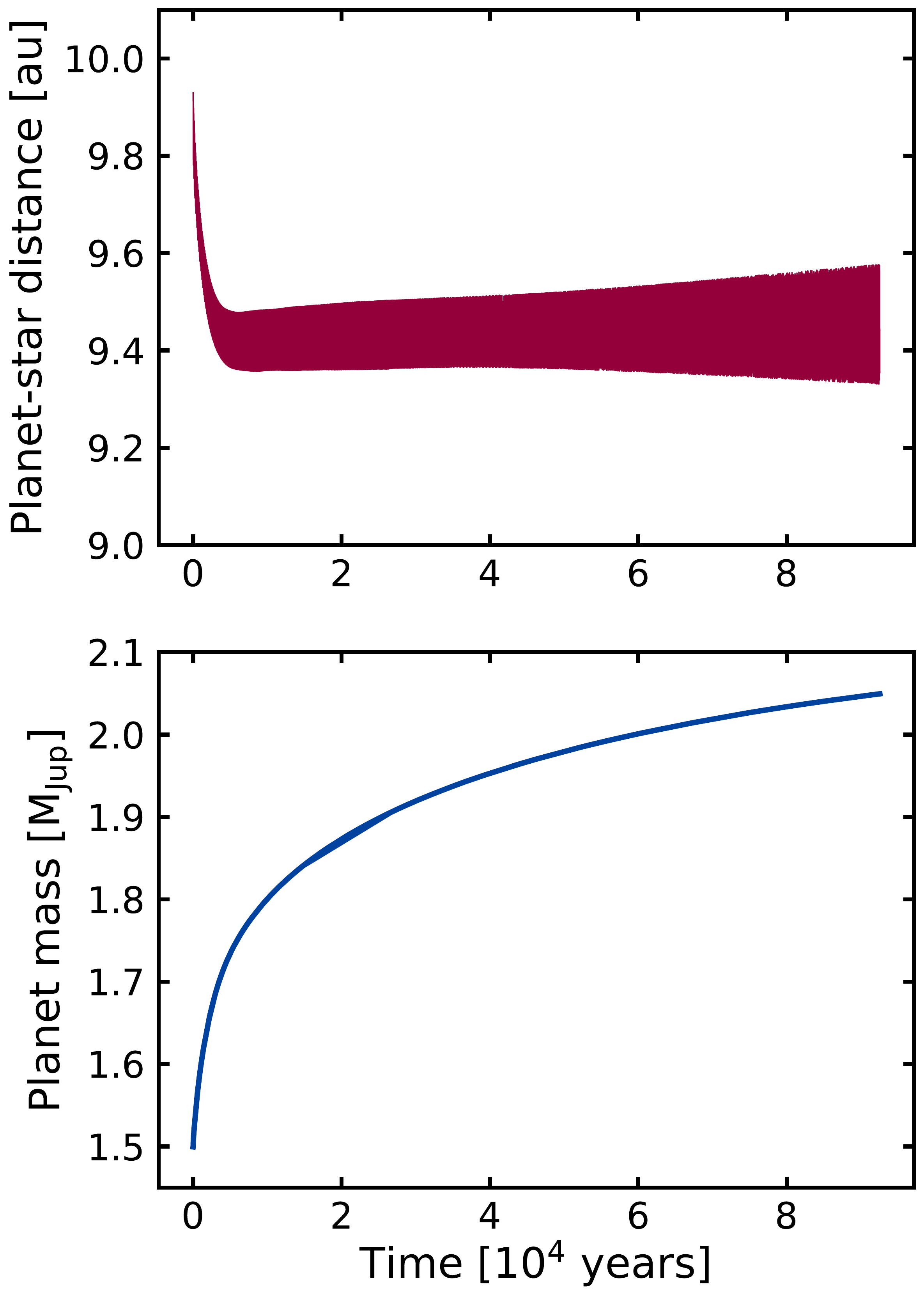}
      \caption{Time evolution of planet-star distance (\textit{top panel}) and planet mass (\textit{bottom panel}) from the simulation with an initial planet mass of $1.5 \, \mathrm{M}_{\mathrm{Jup}}$.}
         \label{fig:planet_evolution}
   \end{figure}

In SPH simulations with \mbox{\textsc{Phantom}}, the planet accretion of gas and dust contained in the disk is self-consistently computed, and the planet migration is directly calculated from the disk gravitational torque, without the need for any prescriptions \citep{2018PASA...35...31P}. In Fig.~\ref{fig:planet_evolution}, we report how planet-star distance and planet mass change with time, in the case of the simulation with an initial planet mass of $1.5 \, \mathrm{M}_{\mathrm{Jup}}$. In both plots, a fast transient is evident in the first $\SI{e4}{\mathrm{yrs}}$, corresponding to the phase of gap opening. After that, the planet stabilizes in a slow type II migration with a smaller increase in mass. Therefore, we can notice that at $\SI{4e4}{\mathrm{yrs}}$, the evolution time at which most of our analyses are performed, the simulation has overcome the initial transient and is in a slowly evolving phase.

From the upper plot in Fig.~\ref{fig:planet_evolution}, a spread in the planet-star distance is visible, indicating an elliptical orbit of the planet in our simulation. Knowing the periastron $r_\mathrm{p}$ and apoastron $r_\mathrm{a}$, we can calculate the planet orbital eccentricity:
\begin{equation}
    e=\frac{r_\mathrm{a}-r_\mathrm{p}}{r_\mathrm{a}+r_\mathrm{p}} \, ,
\end{equation}
obtaining a value $e\approx 0.01$ after $\SI{8e4}{\mathrm{yrs}}$.

\section{Dust mass rescaling}
\label{appendix:dust_mass_rescaling}

   \begin{figure}[h]
   \centering
   \includegraphics[width=\hsize]{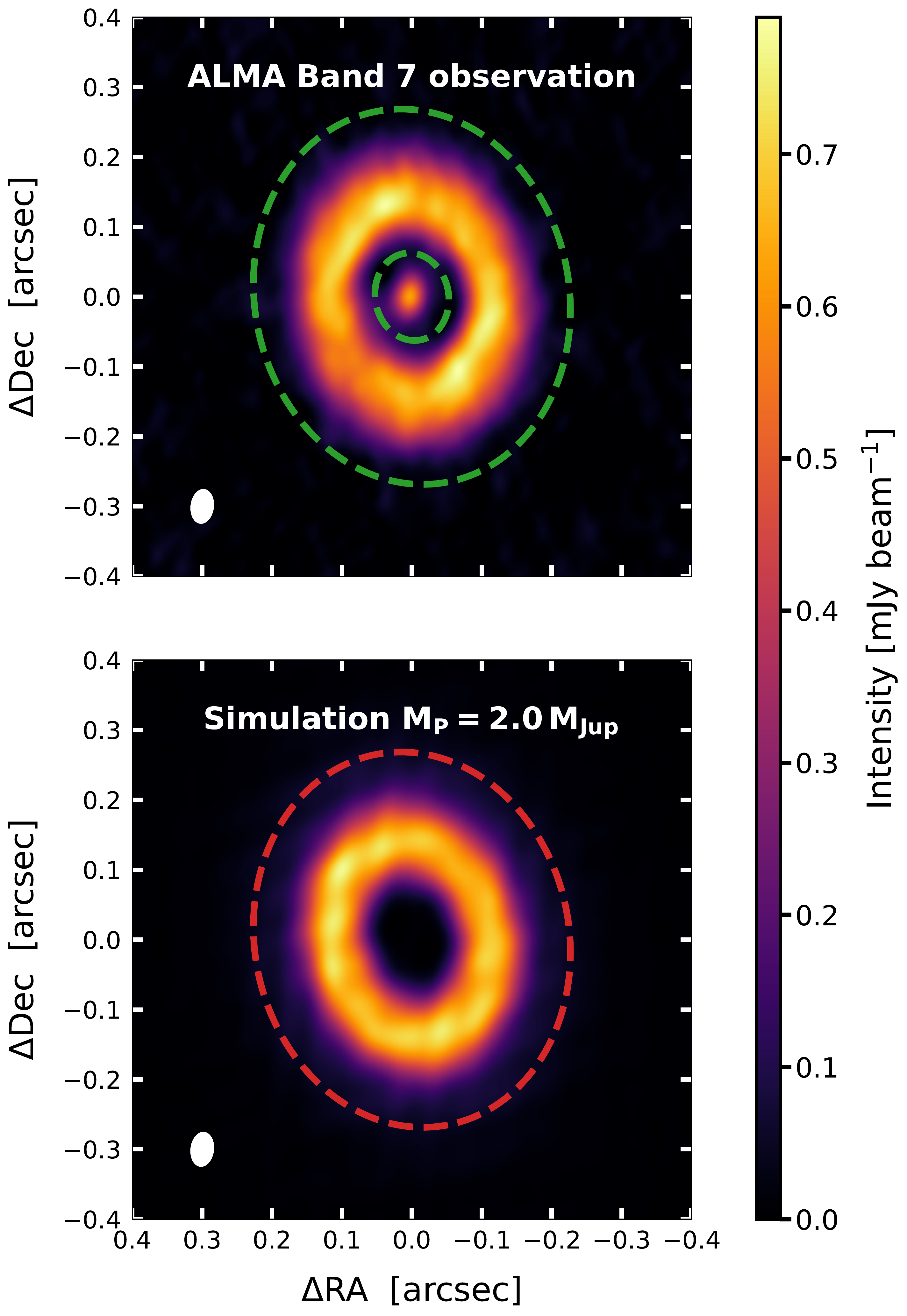}
      \caption{Visual representation of the areas considered for the flux calculation to rescale the total dust masses in our simulations. We take as a reference the flux from the ALMA observation in Band 7, which is contained in the region between the two dashed green ellipses in the \textit{top panel}. Then, we rescale the total dust mass in our simulations so that, from the synthetic images in Band 7, the flux within the dashed red ellipse shown in the \textit{bottom panel} matches the reference flux from the observation. In the \textit{bottom panel}, we display, as an example, the image obtained from the simulation where the planet reaches a mass of $2.0 \, \mathrm{M}_{\mathrm{Jup}}$ after $\SI{4e4}{\mathrm{yrs}}$.
              }
         \label{fig:areas_dust_mass_rescaling}
   \end{figure}

For obtaining the final synthetic dust continuum images from our simulations, we rescale the total dust masses in our models by a constant factor to match the observed flux. In our simulations, the dust-to-gas mass ratio always remains~$\ll 1$. This condition allows this method to be applied without altering the disk dynamics since the dust back-reaction onto the gas remains insignificant. As explained in Sect.~\ref{sect:focus_external_disk}, in our modeling, we focus on reproducing the observed external dust ring. For this reason, we take as reference flux the one emitted only by the external ring in the ALMA Band 7 observation. In particular, we consider the area contained in the two green dashed ellipses depicted in the top panel of Fig.~\ref{fig:areas_dust_mass_rescaling}. The outer ellipse is positioned in the image center, has a semimajor axis of $0.27\arcsec$, a semiminor axis of $0.22\arcsec$, and an inclination of $11^\circ$; for the inner ellipse, we only change the semiaxis length to $0.13\arcsec$ and $0.10\arcsec$. The flux contained in this elliptical annulus is $\SI{29.0}{\mathrm{mJy}}$. In our synthetic images, without an inner disk, we consider the flux contained within the red dashed ellipse in the bottom panel of Fig.~\ref{fig:areas_dust_mass_rescaling}, which has the same parameters as the outer ellipse used for the observation image. We accept the total dust masses that lead to fluxes from our synthetic images from $28.5$ to $\SI{29.5}{\mathrm{mJy}}$.

\section{Dust emission profiles at longer evolution times}
\label{appendix:dust_emission_profiles_8e4yrs}

\begin{figure*}[h!]
   \centering
   \includegraphics[width=0.72\textwidth]{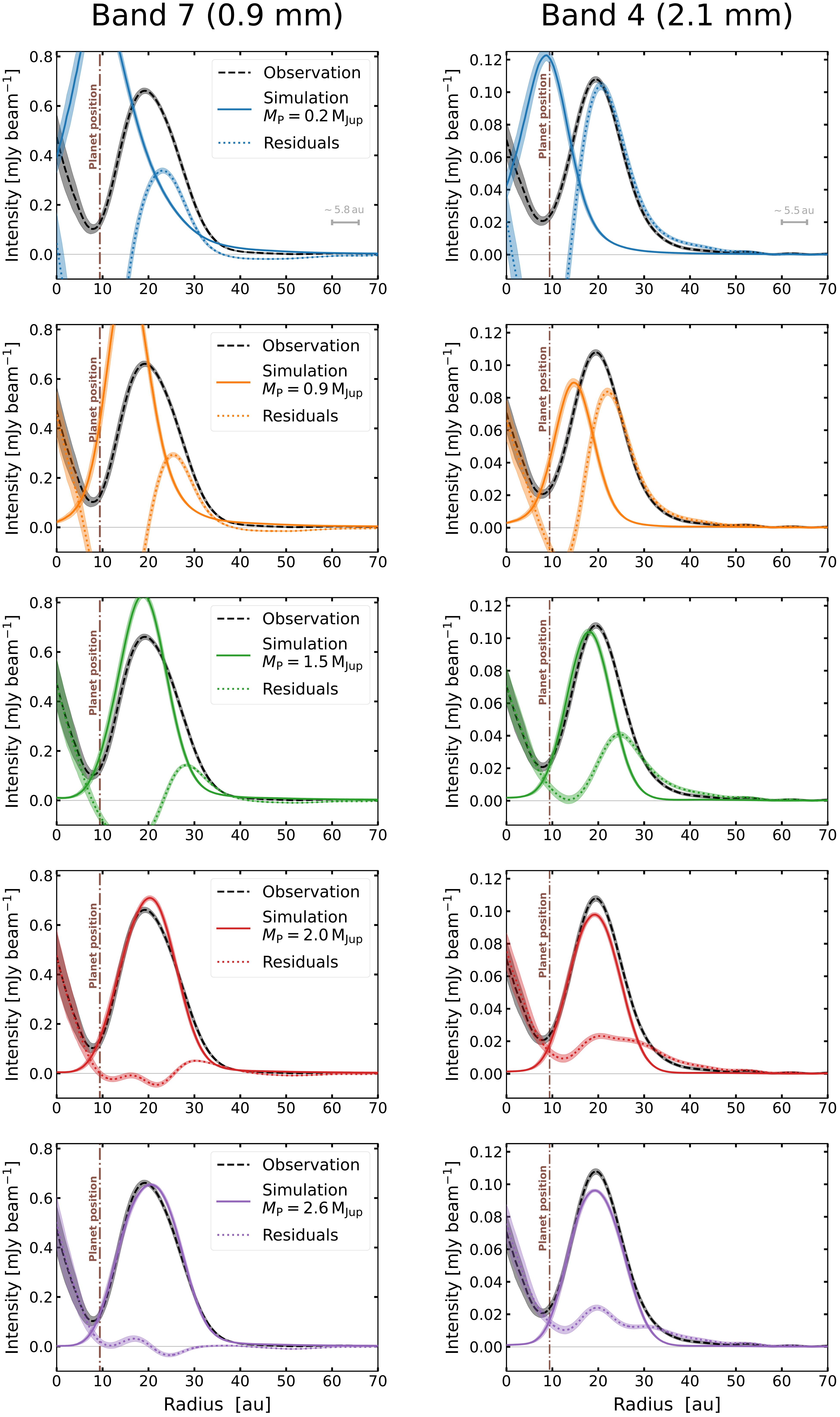}
      \caption{As in Fig.~\ref{fig:mod&res_profiles}, but for the case of simulations evolved for $\SI{8e4}{\mathrm{yrs}}$.}
         \label{fig:mod&res_profiles_8e4yrs}
   \end{figure*}

Fig.~\ref{fig:mod&res_profiles_8e4yrs} shows the dust emission radial profiles in Band~7 and Band~4 of each simulation after an evolution time of \SI{8e4}{\mathrm{yrs}}. Compared to the results after \SI{4e4}{\mathrm{yrs}} (Fig.~\ref{fig:mod&res_profiles}), dust grains had more time to undergo radial drift and possibly accrete onto the star. 

We still note that a final planet mass of at least $2.0 \, \mathrm{M}_{\mathrm{Jup}}$ best reproduces the observations. The slight intensity excess just outside the dust ring has been removed by radial drift. The small mismatch in the peak flux for the simulation with $M_{\mathrm{P}} = 2.0 \, \mathrm{M}_{\mathrm{Jup}}$ and $M_{\mathrm{P}} = 2.6 \, \mathrm{M}_{\mathrm{Jup}}$ in Band~4 is due to the fact that, at this time, the bigger millimeter-sized grains with faster radial drift accreted more onto the central star, thus reducing the continuum emission at this longer wavelength.

\section{Cloud absorption}
\label{appendix:cloud_abs}

   \begin{figure}[]
   \centering
   \includegraphics[width=\hsize]{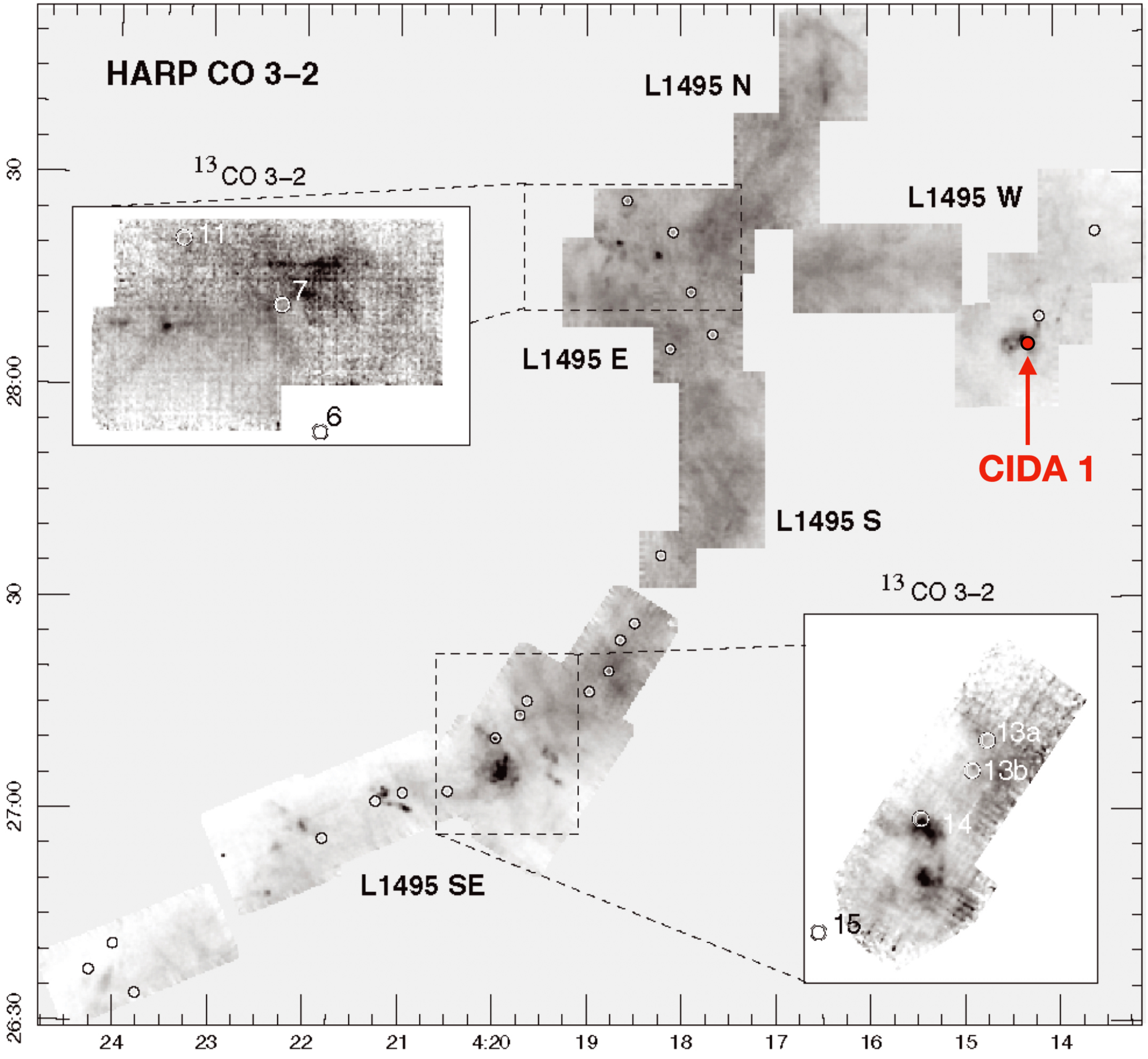}
      \caption{Map of the Taurus molecular cloud obtained by the James Clerk Maxwell Telescope using the $^{12}$CO ($J=3{-}2$) emission (integrated between $2.0$ and \SI{11.0}{\km\,\s^{-1}}). Areas also mapped  with $^{13}$CO ($J=3{-}2$) emission (integrated between $4.9$ and \SI{9.1}{\km\,\s^{-1}}) are indicated by rectangular boxes. We calculated the optical depth property in the subregion L1495~E, whereas CIDA~1 is located in the region L1495~W. The image is adapted from right-hand panel of Fig. 1  in \citet{2010MNRAS.405..759D}. 
              }
         \label{fig:Davis2010_Taurus_map}
   \end{figure}

We aim here to qualitatively estimate the cloud absorption affecting the observed gas channel maps. \citet{2010MNRAS.405..759D} mapped the $^{12}$CO   emission at the transition $J=3{-}2$ in the region of the Taurus molecular cloud, using the Heterodyne Array Receiver Programme (HARP) on the James Clerk Maxwell Telescope. Figure~\ref{fig:Davis2010_Taurus_map} shows the $^{12}$CO map of Taurus that they obtained. The authors divided the region into five different subregions, and in a couple of them, namely L1595~E and L1495~SE, they also collected the $^{13}$CO $J=3{-}2$ emission. CIDA~1 is located in the subregion L1495~W. Since in this area the $^{13}$CO data are missing, we consider the nearby subregion L1595~E for our estimates.

\begin{figure}[]
   \centering
   \includegraphics[width=\hsize]{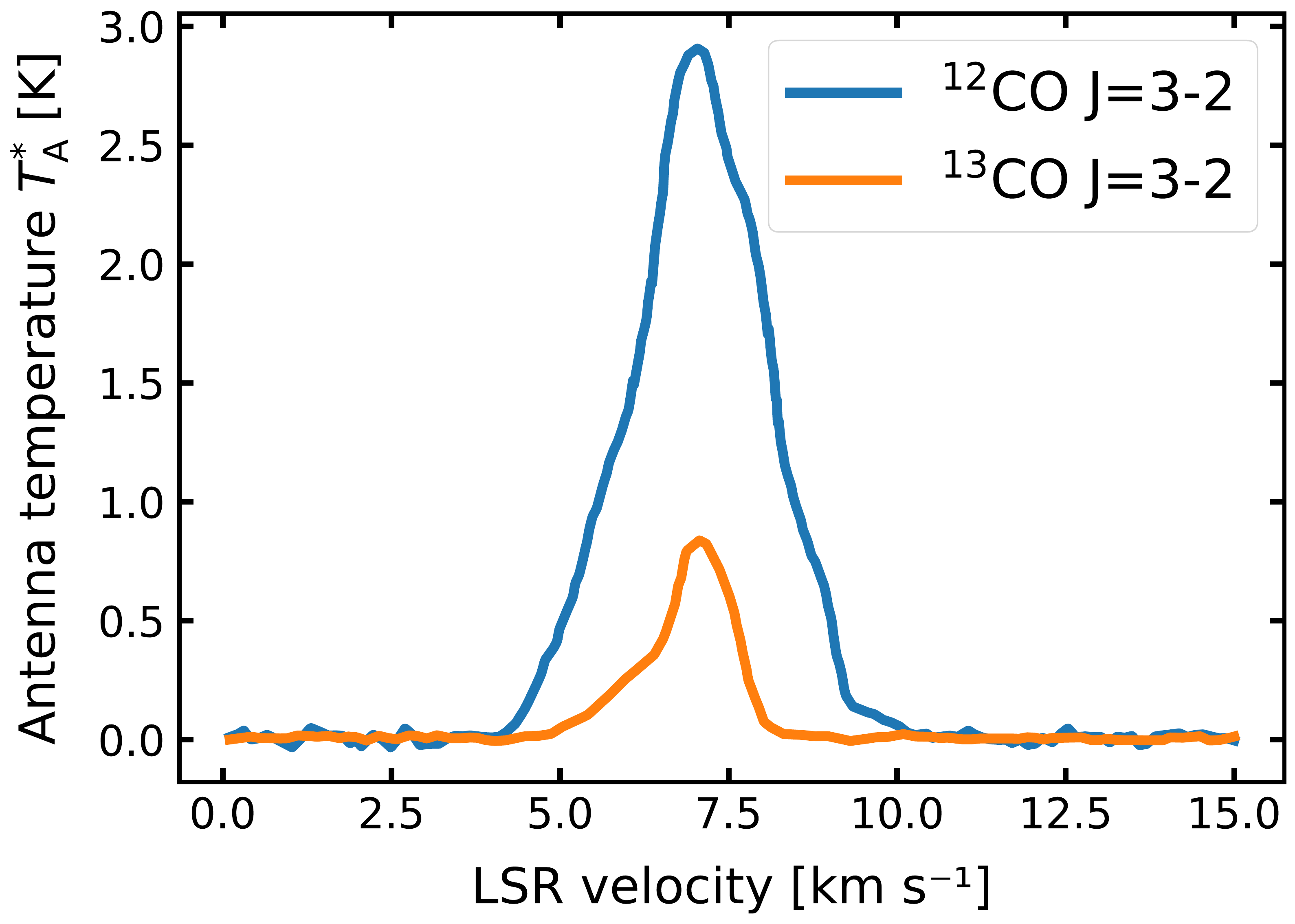}
      \caption{$^{12}$CO ($J=3{-}2$) and $^{13}$CO ($J=3{-}2$) spectra with respect to the local standard of rest (LSR) in the Taurus subregion L1495~E. The spectra are averaged over the entire extent of the subregion. Data are from \citet{2010MNRAS.405..759D}. 
              }
         \label{fig:Davis2010_ant_temp}
   \end{figure}

\begin{figure}[]
   \centering
   \includegraphics[width=\hsize]{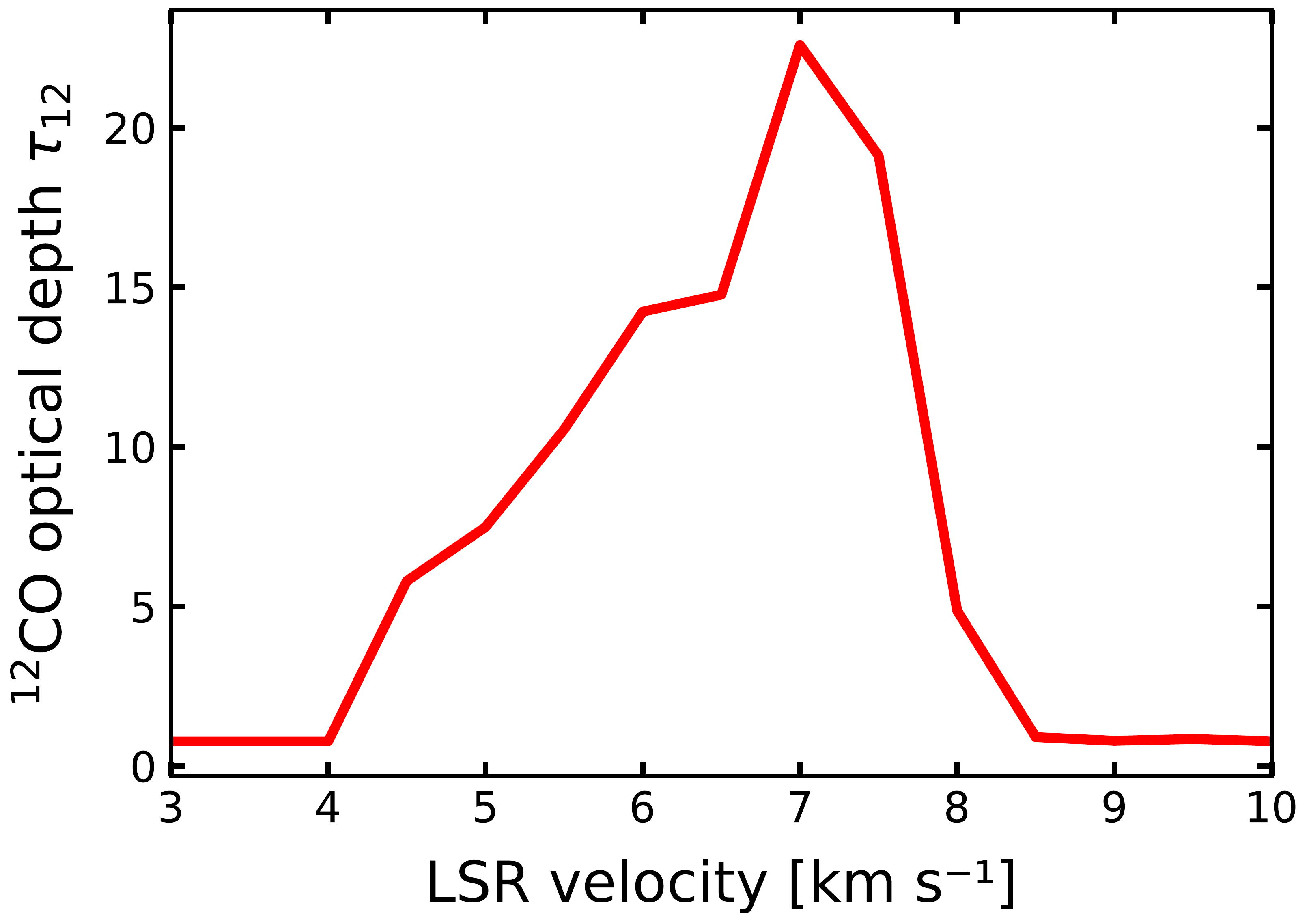}
      \caption{$^{12}$CO ($J=3{-}2$) optical depth in the Taurus subregion L1495~E, calculated from the spectra in Fig~\ref{fig:Davis2010_ant_temp} using Eq.~\ref{eq:Davis2020_optical_depth}. We focus only on the velocity interval where line emission is detected. Bins replicate the central velocity and resolution of the observed $^{12}$CO channel maps. 
              }
         \label{fig:Davis2010_12CO_opt_depth}
   \end{figure}

Figure~\ref{fig:Davis2010_ant_temp} reports the $^{12}$CO and $^{13}$CO spectra of the antenna temperature ($T^{*}_{A}$). As stated in \citet{2010MNRAS.405..759D}, assuming the abundance ratio  $[^{12}\mathrm{CO}/^{13}\mathrm{CO}] = 70$, the $^{12}$CO optical depth $\tau_{12}$ can be calculated as:
\begin{equation}
    \mathlarger{\dfrac{\int T^*_{\mathrm{A}} (^{12}\mathrm{CO}) \, \mathrm{d}\varv}{\int T^*_{\mathrm{A}} (^{13}\mathrm{CO}) \, \mathrm{d}\varv} = \dfrac{1-\mathrm{e^{-\tau_{12}}}}{1-\mathrm{e^{-\tau_{13}}} }} \, ,
    \label{eq:Davis2020_optical_depth}
\end{equation}
where $\tau_{13} = \tau_{12}/70$ is the $^{13}$CO optical depth. We integrate the antenna temperatures along velocity intervals with the same centers and width of the observed $^{12}$CO channel maps (top panels in Fig.~\ref{fig:obs&sim_channels_12CO}). The resulting $^{12}$CO optical depth is presented in Fig.~\ref{fig:Davis2010_12CO_opt_depth}. The morphology of the optical depth spectrum follows approximately the cloud absorption behavior on the $^{12}$CO observation, being strongest at ${\sim}6.0{-}\SI{7.0}{\km\per\s}$ while fading at ${\sim}\SI{4.0}{\km\per\s}$ and ${\sim}\SI{8.5}{\km\per\s}$, where there is a better match between the intensity profiles from our model and the ones from the observation (Fig.~\ref{fig:12CO_channels_comparison}). Nonetheless, it must be stressed that our estimates on the $^{12}$CO spectrum are based on data from a region nearby the one hosting CIDA~1 and, furthermore, these data refer to the total emission of the cloud along the line of sight, while the position of CIDA~1 in this direction, and so the gas column density between us and the source, is unknown.

\end{appendix}

\end{document}